\documentclass[journal]{IEEEtran}

\usepackage{cite}
\usepackage{amsmath,amssymb}
\usepackage{booktabs}
\usepackage{array}
\usepackage{multirow}
\usepackage[pdftex]{graphicx}
\graphicspath{{figures/}}
\usepackage[caption=false,font=footnotesize]{subfig}
\usepackage[final]{changes}
\usepackage{microtype}
\usepackage{tikz}
\usetikzlibrary{arrows.meta,positioning,calc,fit,shapes.geometric}
\begin{document}
	
	\title{Flexible Analog Self-Interference Cancellation for Waveform-Robust Monostatic ISAC}
	
	\author{Anh Tuyen Le,~\IEEEmembership{Member,~IEEE}, Xiaojing Huang,~\IEEEmembership{Senior Member,~IEEE},\\
		J. Andrew Zhang,~\IEEEmembership{Senior Member,~IEEE}, Peiyuan Qin,~\IEEEmembership{Senior Member,~IEEE},  Le Chung Tran,\\~\IEEEmembership{Senior Member,~IEEE},  Nhan Thanh Nguyen,~\IEEEmembership{Senior Member,~IEEE}, and Y. Jay Guo,~\IEEEmembership{Fellow,~IEEE}%
        \thanks{This work has been submitted to the IEEE for possible publication.
            Copyright may be transferred without notice, after which this version may
            no longer be accessible.}%
		\thanks{A. T. Le, X. Huang, A. Zhang, and Y. Jay Guo are with Faculty of Engineering and IT, University of Technology Sydney, Ultimo, NSW, 2007, Australia (emails: anhtuyen.le@uts.edu.au, xiaojing.huang@uts.edu.au, andrew.zhang@uts.edu.au, peiyuan.qin@uts.edu.au, and jay.guo@uts.edu.au). L. C. Tran is with Faculty of Engineering and Information Sciences, University of Wollongong, Wollongong, NSW, 2522 Australia (email: lctran@uow.edu.au). N. T. Nguyen is with the Centre for Wireless Communications, University of Oulu, Oulu, Finland (email: nhan.nguyen@oulu.fi). A preliminary version of this work, introducing the FASIC structure with a fixed-step adaptation, was presented at the IEEE Workshop on Signal Processing Advances in Wireless Communications (SPAWC), 2025.}}
	
	\markboth{Submitted to IEEE Transactions on Wireless Communications}%
	{Le \emph{et al.}: Flexible Analog Self-Interference Cancellation for Waveform-Robust Monostatic ISAC}
	
	\maketitle
	
	\begin{abstract}
		Monostatic integrated sensing and communication (ISAC) requires in-band full-duplex operation, where self-interference cancellation (SIC) must suppress the strong transmit leakage while preserving weak sensing echoes. In the analog domain, this task is complicated by transmitter impairments, which generate conjugate and nonlinear SI components, and by waveforms with widely different statistics. This paper develops a flexible analog self-interference cancellation (FASIC) framework for monostatic ISAC based on the widely linear normalized least-mean-square (WL-NLMS) algorithm. FASIC generates the cancellation signal in digital baseband, subtracts it in the radio-frequency domain, and enables flexible sensing-range control via a configurable cancellation span. Its normalized adaptation remains stable across waveforms with substantially different peak-to-average power ratios (PAPRs), for which fixed-step adaptation can become unstable. We further relate transmit-waveform statistics to analog SIC performance and predict the waveform-dependent cancellation floor. Simulations validate the analysis and demonstrate stable analog SIC of approximately $30$--$34$~dB, with the low-PAPR waveforms achieving a $3.3$~dB deeper floor owing to their smaller high-order amplitude moments. Near-field static reflections within the cancellation span are suppressed while delayed and Doppler-shifted echoes are preserved. These results establish FASIC as a waveform-robust analog SIC solution and identify waveform statistics as a key consideration for sixth-generation (6G) waveform selection.
		
	\end{abstract}
	
	\begin{IEEEkeywords}
		Monostatic ISAC, in-band full-duplex, flexible analog self-interference cancellation (FASIC), widely linear NLMS, OEMP basis, sensing preservation, channel impulse response estimation, range-Doppler imaging, PAPR, OFDM, OTFS, chirped DFT-s-OFDM.
	\end{IEEEkeywords}
	
	\IEEEpeerreviewmaketitle
	
	\section{Introduction}
	\IEEEPARstart{I}{ntegrated} sensing and communication (ISAC) is widely recognized as a key technology for sixth-generation (6G) wireless networks because it enables sensing and data transmission to share spectrum, hardware, and signal-processing resources on a common radio platform \cite{ZhangJSTSP2021}. Among the proposed architectures, the monostatic configuration is particularly attractive for autonomous systems, extended reality, and smart manufacturing. By co-locating the transmitter and receiver, it provides direct sensing observability without the stringent clock synchronization required in bistatic deployments \cite{LiuJSAC2022}. However, this architecture requires in-band full-duplex (IBFD) operation, where strong self-interference (SI) becomes the dominant impairment. In monostatic ISAC, SI comprises the direct transmitter-to-receiver leakage and reflections from nearby objects. These components are typically orders of magnitude stronger than the desired sensing echoes and must therefore be substantially suppressed. Practical SI suppression consequently relies on a combination of antenna isolation, analog radio-frequency (RF) cancellation, and digital cancellation \cite{SmidaJSAC2023,SmidaSPM2024}.
	
	Analog self-interference cancellation (SIC) plays a fundamentally different role in monostatic ISAC than in conventional IBFD communication. Rather than maximizing the cancellation depth alone, the analog canceler must selectively suppress the direct leakage and nearby SI components while preserving delayed echoes that contain useful sensing information \cite{LeTWC2026}. Consequently, its temporal cancellation span becomes as important as its cancellation depth. An analog canceler with an uncontrolled delay span can inadvertently remove weak static or quasi-static sensing echoes together with the SI, degrading sensing performance.

	Although monostatic ISAC depend on effective SIC, much of the existing literature treats SIC as an enabling assumption rather than part of the waveform and receiver design. Holistic trade-off analyses often assume sufficient antenna isolation for the sensing receiver to operate without explicitly modeling SI \cite{KeskinTWC2025}. Full-duplex ISAC waveform studies commonly represent residual SI as an aggregate Gaussian disturbance and report the suppression level required for sensing \cite{XiaoJSAC2022}. Other designs avoid simultaneous leakage by separating radar and communication transmissions into different time slots of the transmission frame, which requires redesigned air-interface resources \cite{ArousIoTJ2026}. Payload-based sensing schemes likewise assume that the echo of the transmitted frame is already available at the monostatic receiver \cite{LiCSM2026,WangTCCN2026}. Even when SIC is considered explicitly, the focus is usually on spatial processing or digital cancellation after the receiver front-end \cite{HeJSAC2023,DuanTVT2024}. Experimental full-duplex orthogonal frequency-division multiplexing (OFDM) radar studies instead show that practical systems require both RF and digital cancellation to suppress SI before and after digitization \cite{BarnetoTMTT2019}.
	
	Most analog RF cancelers, however, have been developed for IBFD communication, where maximizing cancellation depth is the primary objective. Representative designs include multi-tap, frequency-domain, photonic, and adaptive least-mean-square (LMS) based architectures \cite{KwakTWC2023,HongTWC2023,ShiJLT2023,LeTWC2026}. These designs do not directly provide the delay-span selectivity required in monostatic ISAC, where nearby SI should be removed while delayed sensing echoes remain available. This gap motivates an analog SIC architecture that combines front-end protection, impairment-aware SI reconstruction, and controllable delay-span selectivity.
	
	The transmitted waveform also affects analog SIC. Since the waveform passes through the same impaired RF chain that generates the SI, its statistics shape the nonlinear and conjugate components presented to the adaptive canceler. Two waveforms with similar communication and sensing performance can therefore impose different cancellation burdens under the same hardware conditions. Despite this coupling, waveform design and SIC have largely evolved separately. ISAC waveform studies typically emphasize spectral efficiency, ambiguity behavior, and delay--Doppler resolution \cite{SturmPIEEE2011,ZhangJSTSP2021,GaudioTWC2020,WeiWCM2021}. SIC studies usually focus on analog cancellation, nonlinear SI modeling, I/Q imbalance, and adaptive reconstruction for a fixed transmit waveform \cite{SmidaJSAC2023,SmidaSPM2024,FASICSPAWC2025}. As a result, the impact of waveform statistics on analog SIC remains insufficiently characterized. Furthermore, very few studies have addressed analog SIC as an integral part of the monostatic ISAC transceiver design.
	
	\par To address these challenges, we develop the Flexible Analog Self-Interference Cancellation (FASIC) framework for monostatic ISAC. FASIC generates the cancellation signal digitally and subtracts it in the RF domain before the receiver front-end. This structure provides front-end protection, reconstructs direct, conjugate, and nonlinear SI components, and allows the cancellation span, namely the range of echo delays that the canceler can absorb, to be configured. A preliminary version of FASIC with fixed-step widely linear LMS (WL-LMS) adaptation was presented in \cite{FASICSPAWC2025}. Making such an analog canceler robust to the transmit waveform, however, raises a challenge beyond that preliminary design. The cancellation signal is reconstructed adaptively from nonlinear functions of the transmit samples, so the stability and accuracy of the adaptation depend on the waveform statistics. In particular, an adaptation step size that is stable for a low peak-to-average power ratio (PAPR) waveform can diverge for high-PAPR multicarrier signals under the same canceler and basis representation. This paper addresses this challenge with a widely linear normalized LMS (WL-NLMS) update, which scales the step size by the instantaneous energy of the reconstruction inputs so that structurally different waveforms are handled with a single setting. Built on this update, the paper completes the waveform-robust design and analyzes how transmit-waveform statistics determine analog SIC performance, focusing on second-order complementary correlation and the high-order amplitude moments associated with PAPR. We perform extensive simulations to evaluate FASIC with four representative candidates: single-carrier (SC) transmission, OFDM, orthogonal time frequency space (OTFS) modulation, and chirped discrete Fourier transform (DFT) spread OFDM (DFT-s-OFDM), under a common cancellation architecture, power-amplifier (PA) model, and passive-suppression level. Under this common configuration, FASIC with WL-NLMS achieves stable analog SIC of approximately $30$--$34$~dB across all four candidates, and the low-PAPR waveforms achieve about $3.3$~dB deeper SIC, consistent with the analysis.
	
	The main contributions of this paper are summarized as follows:
	
	\begin{itemize}
		\item \textbf{Waveform-robust FASIC with WL-NLMS:} We develop FASIC with the WL-NLMS update for monostatic ISAC, in which the adaptation step is automatically scaled by the instantaneous energy of the reconstruction inputs. This keeps the analog cancellation stable across structurally different waveform candidates with a single normalized step.
		\item \textbf{Waveform-statistical analysis:} We analyze how the transmit-waveform statistics affect the widely linear canceler and separate two mechanisms: second-order impropriety, which the widely linear structure accommodates, and PAPR-driven high-order amplitude moments, which determine the nonlinear SI power that the finite cancellation basis cannot fully reconstruct. The analysis establishes the stability and first-order steady-state behavior of the normalized update.
		\item \textbf{SIC-depth characterization:} We establish an explicit relationship between the residual SIC and the normalized amplitude moments $\gamma_p$, showing that the highest retained moment $\gamma_P$ governs the waveform-dependent cancellation floor under finite-order and finite-tap reconstruction.
		\item \textbf{Sensing preservation and span flexibility:} We demonstrate sensing preservation across all four waveforms through channel impulse response and time-of-arrival estimation, as well as range-Doppler imaging. We also show that the configurable cancellation span enables delay-selective suppression of near-field static clutter while preserving delayed or Doppler-shifted sensing echoes.
	\end{itemize}
	
	The rest of this paper is organized as follows. Section~II presents the monostatic ISAC system model and the nonlinear SI model. Section~III develops the proposed FASIC structure with the WL-NLMS adaptation and analyzes the canceler's performance from the viewpoint of waveform statistics, characterizing its stability and steady-state behavior. Section~IV introduces the candidate waveforms and develops testable predictions for waveform-dependent SIC. Section~V presents the numerical results, covering the cancellation comparison across waveforms, the waveform-statistics characterization, the sensing-preservation evaluation, and the demonstration of the flexible cancellation span. Section~VI concludes the paper.
	
	\emph{Notation:} Boldface lowercase and uppercase letters denote column vectors and matrices, respectively. The superscripts $(\cdot)^{*}$, $(\cdot)^{T}$, and $(\cdot)^{H}$ denote complex conjugation, transpose, and Hermitian transpose, where conjugation applies elementwise to vectors and matrices. $\mathbb{E}\{\cdot\}$ denotes expectation, $\mathrm{tr}(\cdot)$ and $\lambda_{\max}(\cdot)$ denote the trace and the largest eigenvalue of a matrix, $\|\cdot\|$ is the Euclidean norm, and $*$ denotes linear convolution.
	
	\section{System Model}
	\begin{figure}[!t]
		\centering
		\includegraphics[width=\columnwidth]{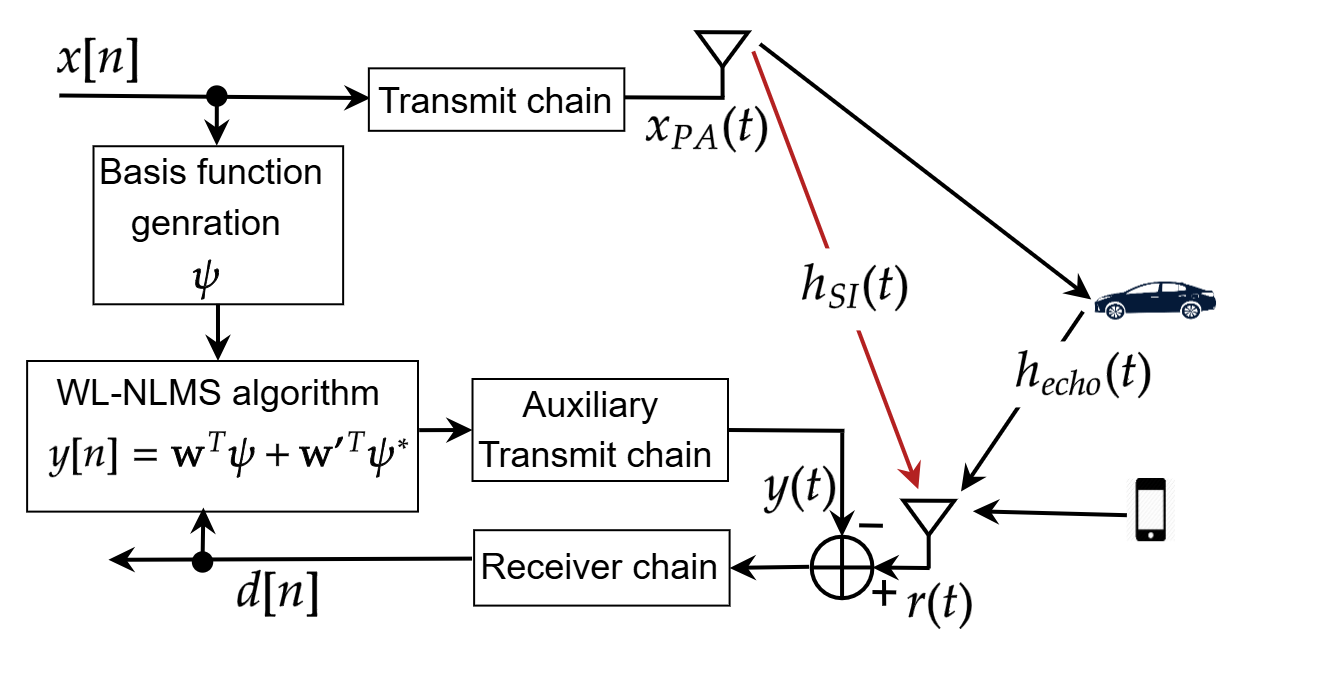}
		\caption{Monostatic ISAC transceiver with the FASIC canceler.}
		\label{fig:system_architecture}
	\end{figure}
	\subsection{Transmit Signal and RF Front-End Model}
	Consider the monostatic ISAC transceiver in Fig.~\ref{fig:system_architecture}. Let $x[n]$ and $x(t)$ denote the discrete-time digital and continuous-time analog baseband transmit signals, respectively, where the symbol period is $T_s$. The waveform may be SC, OFDM, OTFS, or chirped DFT-s-OFDM, depending on the communication and sensing requirements. After digital-to-analog conversion, $x(t)$ passes through the transmitter RF chain, including the I/Q up-conversion mixer and the PA.
	\par Hardware imperfections in the transmitter mixer introduce frequency-dependent I/Q imbalance (IQI) \cite{LeTVT2022IQI}. Let $h_I(t)$ and $h_Q(t)$ denote the real-valued impulse responses of the I- and Q-branch low-pass filters. Let $g_t$ and $\phi_t$ denote the amplitude and phase mismatches between the two branches. The effective direct and conjugate IQI responses, $u_t(t)$ and $v_t(t)$, are
	\begin{align}
		u_t(t) &= \tfrac{1}{2}\!\left[\,h_I(t) + g_t\,h_Q(t)\,e^{j\phi_t}\right],\\
		v_t(t) &= \tfrac{1}{2}\!\left[\,h_I(t) - g_t\,h_Q(t)\,e^{j\phi_t}\right].
		\label{eq:iqi_responses}
	\end{align}
	The impaired mixer output is therefore modeled as the sum of a direct component and a conjugate leakage component,
	\begin{align}
		x_{\mathrm{IQ}}(t) = u_t(t)*x(t) + v_t(t)*x^{*}(t).
		\label{eq:iqi}
	\end{align}
	
	To improve power efficiency and extend the sensing range, the PA is often operated close to saturation. The resulting nonlinear distortion is modeled by the baseband-equivalent parallel Hammerstein model \cite{DingTCOM2004,MorganTSP2006},
	\begin{align}
		x_{\mathrm{PA}}(t) &= \sum_{\substack{p=1\\p\,\mathrm{odd}}}^{P} \alpha_p(t)*\!\left(x_{\mathrm{IQ}}(t)\,|x_{\mathrm{IQ}}(t)|^{p-1}\right),
		\label{eq:pa_model}
	\end{align}
	where $\alpha_p(t)$ is the $p$-th-order complex impulse response capturing PA memory, and $P$ is the highest nonlinear order.
	In discrete time, the $p$-th-order response is an $M_{\mathrm{PA}}$-tap filter $a_{p,m}\triangleq\alpha_p(mT_s)$, $m=0,\ldots,M_{\mathrm{PA}}-1$, with $M_{\mathrm{PA}}$ being the PA memory depth. We use $\alpha_p(t)$ and $\{a_{p,m}\}$ interchangeably.
	
	In monostatic ISAC, the receiver observes the SI, sensing echoes, a possible uplink communication signal, and thermal noise. The receiver input is decomposed as
	\begin{align}
		r(t) &= s_{\mathrm{SI}}(t) + s_{\mathrm{echo}}(t) + s_{\mathrm{far}}(t) + v(t),
		\label{eq:rx_decomp}
	\end{align}
	where $s_{\mathrm{SI}}(t)$ denotes the SI signal, $s_{\mathrm{echo}}(t)$ represents the sensing echo reflected from surrounding targets, $s_{\mathrm{far}}(t)$ is the uplink communication signal transmitted by the far-end user, and $v(t)$ is the receiver noise.
	
	\subsection{Nonlinear SI Model and Basis Expansion}
	Let $h_{\mathrm{SI}}(t)$ denote the SI channel between the transmitter and receiver. The SI signal is the PA output in \eqref{eq:pa_model} propagated through this channel, i.e., $s_{\mathrm{SI}}(t)=h_{\mathrm{SI}}(t)*x_{\mathrm{PA}}(t)$. This cascade can be written in terms of the original transmit signal through the expanded memory polynomial (EMP) model \cite{Huangoemp},
	\begin{align}
		s_{\mathrm{SI}}(t) &\approx \sum_{\substack{p=1\\p\,\mathrm{odd}}}^{P}\sum_{\substack{q=1\\q\,\mathrm{odd}}}^{p}\Big[\alpha_{q,p}(t)*x^{q}(t)|x(t)|^{p-q} \nonumber\\
		&\hspace{2em}+ \beta_{q,p}(t)*x^{*q}(t)|x(t)|^{p-q}\Big] \nonumber\\
		&= \sum_{\substack{p=1\\p\,\mathrm{odd}}}^{P}\sum_{\substack{q=1\\q\,\mathrm{odd}}}^{p}\Big[\alpha_{q,p}(t)*\xi_{q,p}(t) + \beta_{q,p}(t)*\xi^{*}_{q,p}(t)\Big],
		\label{eq:si_basis_model}
	\end{align}
	where the continuous-time EMP basis function is defined as
	\begin{align}
		\xi_{q,p}(t) \triangleq x^{q}(t)\,|x(t)|^{p-q}.
		\label{eq:emp_basis}
	\end{align}
	Here, $\alpha_{q,p}(t)$ and $\beta_{q,p}(t)$ are the effective direct and conjugate responses associated with $\xi_{q,p}(t)$ and $\xi^{*}_{q,p}(t)$. They collect the combined effects of transmitter IQI, PA nonlinearity, and the SI channel.
	
	\section{Proposed FASIC Framework}
	\label{sec:fasic}
	\subsection{FASIC With WL-NLMS Over the OEMP Basis}
	\label{subsec:wllms_update}
	To suppress the SI in \eqref{eq:si_basis_model} before the receiver front-end while controlling the canceled delay range, this paper proposes FASIC over the orthogonal EMP (OEMP) basis with a waveform-robust WL-NLMS adaptation. FASIC generates the cancellation signal in digital baseband and subtracts it in the RF domain through an auxiliary transmit chain, thereby protecting the receiver front-end while allowing the canceled delay range to be configured. Following the OEMP-based formulation in \cite{FASICSPAWC2025}, we represent the SI in \eqref{eq:si_basis_model} over the OEMP basis rather than the raw EMP basis. The raw EMP basis functions are often strongly correlated, especially across higher nonlinear orders, which degrades the conditioning of the adaptive reconstruction. We therefore map the set of functions $\{\xi_{q,p}(t)\}$ to the OEMP basis as \cite{Huangoemp},
	\begin{align}
		\psi_{q,p}(t)=\sqrt{\frac{(\frac{p+q}{2})!}{(\frac{p-q}{2})!}}\!\!\sum_{\substack{p'=q\\ p'\,\mathrm{odd}}}^{p}\!\frac{(-1)^{\frac{p-p'}{2}}\binom{\frac{p-q}{2}}{\frac{p'-q}{2}}}{(\frac{p'+q}{2})!\,\sigma_x^{p'}}\,\xi_{q,p'}(t),
		\label{eq:orth_basis_def}
	\end{align}
	where $\sigma_x^2=\mathbb{E}\{|x(t)|^2\}$ is the transmit-signal power. The transformation in \eqref{eq:orth_basis_def} is triangular and invertible. Hence, the raw EMP and OEMP bases span the same subspace, and the SI model in \eqref{eq:si_basis_model} can be represented in either basis. Under a zero-mean circular complex Gaussian reference model for $x(t)$, the OEMP functions are mutually orthogonal across nonlinear orders,
	\begin{align}
		\mathbb{E}\{\psi_{q,p}(t)\psi^{*}_{q,p'}(t-\tau)\}
		=
		\begin{cases}
			\varphi_{\psi\psi}^{(q,p)}(\tau), & p=p',\\
			0, & p\neq p',
		\end{cases}
		\label{eq:orth_basis_cond}
	\end{align}
	where $\varphi_{\psi\psi}^{(q,p)}(\tau)$ is the autocorrelation function of $\psi_{q,p}(t)$. This orthogonality makes the basis Gram matrix block diagonal across orders under the Gaussian reference model, improving the conditioning and convergence of the adaptive reconstruction \cite{FASICSPAWC2025}.
	\par Using the OEMP basis, FASIC reconstructs the SI as a widely linear combination of the basis functions and their conjugates \cite{PicinbonoTSP1995,MandicGoh2009}. In the equivalent baseband model, the cancellation signal is
	\begin{align}
		y[n] &= \mathbf{w}^{T}[n]\boldsymbol{\psi}[n]
		+ \mathbf{w}'^{T}[n]\boldsymbol{\psi}^{*}[n],
		\label{eq:cancel_matrix}
	\end{align}
	where
	\begin{align}
		\mathbf{w}[n]
		&=
		\big[
		w^{0}_{1,1}[n],\ldots,w^{L-1}_{1,1}[n],
		\ldots,
		w^{0}_{P,P}[n],\ldots,w^{L-1}_{P,P}[n]
		\big]^{T},
		\nonumber\\
		\boldsymbol{\psi}[n]
		&=
		\big[
		\psi_{1,1}(nT),\psi_{1,1}((n{-}1)T),\ldots,
		\nonumber\\
		&\quad\quad
		\psi_{P,P}((n{-}L{+}1)T)
		\big]^{T}
		\in \mathbb{C}^{K}.
		\nonumber
	\end{align}
	Here, \(K\) is the dimension of \(\boldsymbol{\psi}[n]\), \(w^{l}_{q,p}[n]\) is the weighting coefficient of the \(l\)-th delayed basis sample \(\psi_{q,p}((n{-}l)T)\), and \(L\) is the filter length assigned to each basis function. The tap interval \(T\) satisfies \(T\le T_s\) to capture the arbitrary delay of the SI channel. Hence, the \(L\) taps span the delay range from \(0\) to \((L{-}1)T\), which determines the cancellation span. The vector \(\mathbf{w}'[n]\) contains the weighting coefficients of the conjugate branch multiplying \(\boldsymbol{\psi}^{*}[n]\), where the prime is a branch label rather than an operator. The cancellation signal is then up-converted by the auxiliary transmit chain.
	\par After RF cancellation, down-conversion, and digitization, the baseband residual is
	\begin{align}
		d[n] &= k\big(r[n]-y[n]\big),
		\label{eq:residual}
	\end{align}
	where $k$ is the real receive-chain gain representing the combined low-noise-amplifier and receive-mixer gain. The WL-LMS algorithm adapts the coefficients by minimizing $\mathbb{E}\{|d[n]|^2\}$, giving
	\begin{align}
		\mathbf{w}[n{+}1]
		&=
		\mathbf{w}[n]
		+
		\mu\, d[n]\,\boldsymbol{\psi}^{*}[n],
		\nonumber\\
		\mathbf{w}'[n{+}1]
		&=
		\mathbf{w}'[n]
		+
		\mu\, d[n]\,\boldsymbol{\psi}[n],
		\label{eq:wllms}
	\end{align}
	where \(\mu\) denotes the step size. Both recursions follow the stochastic gradient of $|d[n]|^{2}$ with respect to the conjugates of $\mathbf{w}[n]$ and $\mathbf{w}'[n]$ under the transpose-based model \eqref{eq:cancel_matrix}, which is why the error $d[n]$ enters unconjugated.
	
	The fixed-step update in \eqref{eq:wllms} was used in the preliminary FASIC study with SC transmission \cite{FASICSPAWC2025}. Since the nonlinear regressor energy varies strongly across samples for high-PAPR waveforms, we address this sensitivity by the following WL-NLMS update, obtained by replacing the fixed step $\mu$ in \eqref{eq:wllms} with the normalized step
	\begin{align}
		\mu_n=\frac{\mu}{\varepsilon_0+2\|\boldsymbol{\psi}[n]\|^{2}},
		\label{eq:nlms}
	\end{align}
	where the dimensionless constant $\mu$ controls the adaptation rate, $2\|\boldsymbol{\psi}[n]\|^{2}$ is the instantaneous augmented-regressor energy, and $\varepsilon_0$ is a small regularizer.
	
	FASIC also provides flexible delay-span control. Since the cancellation signal is generated digitally, the cancellation span, $(L{-}1)T$, can be adjusted through the filter length $L$ to determine the maximum echo delay absorbed by the canceler. As shown for the linear analog canceler in \cite{LeTWC2026}, the span should cover the entire SI channel while remaining shorter than the delay of the nearest static reflector to be preserved,
	\begin{align}
		\tau_{\mathrm{SI,max}} < (L{-}1)T < \tau_{\min}-2\,T_s,
		\label{eq:span_gate}
	\end{align}
	where $\tau_{\mathrm{SI,max}}$ denotes the maximum SI delay, $\tau_{\min}$ is the delay of the nearest reflector to be preserved, and the $2\,T_s$ margin corresponds to the correlation width of the root-raised-cosine (RRC) transmit pulse. Consequently, static echoes within the cancellation span are suppressed, whereas those beyond the span remain weakly correlated with the cancellation basis and are preserved for sensing. In contrast, moving targets are preserved regardless of the cancellation span because their Doppler-induced phase variation decorrelates the echoes from the static cancellation basis over the adaptation interval.

	The remainder of this section develops the analytical link between transmit-waveform statistics and the performance of the proposed WL-NLMS canceler. Section~\ref{subsec:mean_stability} derives the mean-stability behavior of the augmented recursion, Section~\ref{subsec:emse_misadj} obtains a first-order steady-state excess mean-square error (EMSE) characterization, and Section~\ref{subsec:papr_moments} identifies the high-order amplitude moments that determine the nonlinear SI power presented to a finite cancellation basis.
	
	\subsection{Mean Behavior and Stability of the Adaptive Recursion}
	\label{subsec:mean_stability}
	For the analysis, we decompose the received signal at the canceler input as
	\begin{align}
		r[n] = r_{\mathrm{SI}}[n] + \nu[n],
		\label{eq:r_decomp_analysis}
	\end{align}
	where $r_{\mathrm{SI}}[n]$ denotes the SI component representable in the adopted basis, and the residual term $\nu[n]$ collects all components \emph{not} canceled by the finite-dimensional basis representation, including sensing echoes, far-end communication signals, receiver noise, and any SI mismatch caused by the finite basis order or tap length. The SI signal can be expressed in the augmented model as
	\begin{align}
		r_{\mathrm{SI}}[n] = \underline{\mathbf{w}}_o^T \underline{\boldsymbol{\psi}}[n],
		\label{eq:si_projection}
	\end{align}
	where $\underline{\boldsymbol{\psi}}[n]\triangleq[\boldsymbol{\psi}^{T}[n],\,\boldsymbol{\psi}^{H}[n]]^{T}$ is the augmented basis vector, $\underline{\mathbf{w}}[n]\triangleq[\mathbf{w}^{T}[n],\,\mathbf{w}'^{T}[n]]^{T}$ is the corresponding augmented weighting coefficient vector, and $\underline{\mathbf{w}}_o$ is the optimum weighting coefficient vector associated with the representable SI. Since the receive-chain gain $k$ in \eqref{eq:residual} is a known real scalar, it is set to unity in the analysis without loss of generality. Defining the augmented weight-error vector $\underline{\tilde{\mathbf{w}}}[n]\triangleq\underline{\mathbf{w}}_o-\underline{\mathbf{w}}[n]$ and substituting \eqref{eq:si_projection} into \eqref{eq:residual}, the residual signal becomes
	\begin{align}
		d[n]
		&= r[n]-\underline{\mathbf{w}}^{T}[n]\underline{\boldsymbol{\psi}}[n] \nonumber\\
		&= \underline{\tilde{\mathbf{w}}}^{T}[n]\underline{\boldsymbol{\psi}}[n]+\nu[n].
		\label{eq:residual_analysis}
	\end{align}
	The update \eqref{eq:wllms}, with either the fixed step $\mu$ or the normalized step \eqref{eq:nlms}, can be written in the augmented form as
	\begin{align}
		\underline{\mathbf{w}}[n+1]
		=
		\underline{\mathbf{w}}[n]+\mu[n]\,d[n]\,\underline{\boldsymbol{\psi}}^{*}[n],
		\label{eq:wllms_aug}
	\end{align}
	where $\mu[n]=\mu$ for the fixed-step update and $\mu[n]=\mu_n$ for the proposed WL-NLMS. Subtracting $\underline{\mathbf{w}}_o$ from both sides of \eqref{eq:wllms_aug} and using \eqref{eq:residual_analysis} yield the raw weight-error recursion
	\begin{align}
		\underline{\tilde{\mathbf{w}}}[n+1]
		=
		\Big(\mathbf{I}
		-\mu[n]\,\underline{\boldsymbol{\psi}}^{*}[n]\underline{\boldsymbol{\psi}}^{T}[n]\Big)
		\underline{\tilde{\mathbf{w}}}[n]
		-\mu[n]\,\nu[n]\underline{\boldsymbol{\psi}}^{*}[n].
		\label{eq:raw_weight_error}
	\end{align}
	Denoting the elementwise conjugate of the augmented weight-error vector as $\boldsymbol{\epsilon}[n]\triangleq\big(\underline{\tilde{\mathbf{w}}}[n]\big)^{*}$ and conjugating both sides of \eqref{eq:raw_weight_error} gives
	\begin{align}
		\boldsymbol{\epsilon}[n+1]
		=
		\Big(\mathbf{I}
		-\mu[n]\,\underline{\boldsymbol{\psi}}[n]\underline{\boldsymbol{\psi}}^{H}[n]\Big)
		\boldsymbol{\epsilon}[n]
		-\mu[n]\,\nu^{*}[n]\underline{\boldsymbol{\psi}}[n].
		\label{eq:weight_error_evol}
	\end{align}
	This form is useful because the instantaneous kernel $\underline{\boldsymbol{\psi}}[n]\underline{\boldsymbol{\psi}}^{H}[n]$ is Hermitian positive semidefinite, which allows the mean behavior to be expressed through an augmented correlation matrix. We adopt the following assumptions to make the recursion analytically tractable:
	\begin{itemize}
		\item[\textbf{A1)}] The augmented regressor sequence $\underline{\boldsymbol{\psi}}[n]$ is wide-sense cyclostationary with finite fourth-order moments. Since the transmit signal is linearly modulated with periodic pulse shaping and block structure, $\underline{\boldsymbol{\psi}}[n]$ is cyclostationary rather than strictly stationary. At the small step sizes used for deep SIC, adaptation is slow relative to the cyclostationary period, so the recursion depends on the regressor only through its time-averaged statistics.
		\item[\textbf{A2)}] Under the standard small-step-size approximation, the weight error $\underline{\tilde{\mathbf{w}}}[n]$ (equivalently $\boldsymbol{\epsilon}[n]$) is statistically independent of the instantaneous regressor $\underline{\boldsymbol{\psi}}[n]$, and hence of any measurable function of it, in particular the normalized kernel $\underline{\boldsymbol{\psi}}\underline{\boldsymbol{\psi}}^{H}/(\varepsilon_0+\|\underline{\boldsymbol{\psi}}\|^{2})$.
		\item[\textbf{A3)}] The disturbance $\nu[n]$ is zero mean with variance $\sigma_\nu^2$. It may include receiver noise, far-end communication signals, finite-basis mismatch, and sensing echoes outside the cancellation subspace. For the mean analysis, these components are treated as uncorrelated with the instantaneous augmented regressor. Static sensing echoes satisfy this condition when their delays lie beyond the canceler span, up to the pulse-correlation width. Doppler-shifted echoes can also decorrelate from the static regressors over the adaptation interval.
	\end{itemize}
	Under these assumptions, we define the augmented correlation matrix
	\begin{align}
		\mathbf{R}\triangleq\mathbb{E}\left\{\underline{\boldsymbol{\psi}}[n]\underline{\boldsymbol{\psi}}^{H}[n]\right\}=\begin{bmatrix}\mathbf{C} & \mathbf{P}\\\mathbf{P}^{*} & \mathbf{C}^{*}\end{bmatrix},
		\label{eq:R_aug}
	\end{align}
	where $\mathbf{C}=\mathbb{E}\{\boldsymbol{\psi}[n]\boldsymbol{\psi}^{H}[n]\}$ and $\mathbf{P}=\mathbb{E}\{\boldsymbol{\psi}[n]\boldsymbol{\psi}^{T}[n]\}$ are the covariance and pseudo-correlation matrices, respectively. Throughout this section, $\mathbb{E}\{\cdot\}$ applied to regressor correlations denotes the cyclic mean, that is, ensemble expectation followed by a time-average over the $U$ samples of one symbol period, which renders these correlations time-invariant. For symbol-rate regressors ($U{=}1$) it reduces to ordinary expectation. For the \emph{fixed-step} update, taking the expectation of \eqref{eq:weight_error_evol} and invoking \textbf{A2}--\textbf{A3}, we have
	\begin{align}
		\mathbb{E}\{\boldsymbol{\epsilon}[n+1]\}=\left(\mathbf{I}-\mu\mathbf{R}\right)\mathbb{E}\{\boldsymbol{\epsilon}[n]\}.
		\label{eq:mean_error}
	\end{align}
	Hence, a standard sufficient condition for mean stability is
	\begin{align}
		0<\mu<\frac{2}{\lambda_{\max}(\mathbf{R})}.
		\label{eq:mu_stab}
	\end{align}
	Equation \eqref{eq:mu_stab} is waveform-dependent because $\lambda_{\max}(\mathbf{R})$ scales with the high-order moments of the transmit envelope that drive the nonlinear regressor branches, quantified in Section~\ref{subsec:papr_moments}. A step chosen for a low-PAPR waveform can violate \eqref{eq:mu_stab} for a high-PAPR one. This is the divergence that motivates the proposed update. For the normalized update, the same expectation gives
	\begin{align}
		\mathbb{E}\{\boldsymbol{\epsilon}[n+1]\}&=\left(\mathbf{I}-\mu\mathbf{R}_{\mathrm n}\right)\mathbb{E}\{\boldsymbol{\epsilon}[n]\},
		\nonumber\\
		\mathbf{R}_{\mathrm n}&\triangleq\mathbb{E}\!\left\{\frac{\underline{\boldsymbol{\psi}}[n]\underline{\boldsymbol{\psi}}^{H}[n]}{\varepsilon_0+\|\underline{\boldsymbol{\psi}}[n]\|^{2}}\right\},
		\label{eq:mean_error_nlms}
	\end{align}
	where $\mathbf{R}_{\mathrm n}$ is the normalized augmented correlation matrix. Since $\mathbf{R}_{\mathrm n}\succeq\mathbf{0}$ and $\mathrm{tr}(\mathbf{R}_{\mathrm n})=\mathbb{E}\{\|\underline{\boldsymbol{\psi}}\|^2/(\varepsilon_0+\|\underline{\boldsymbol{\psi}}\|^2)\}<1$, every eigenvalue of $\mathbf{R}_{\mathrm n}$ lies in $[0,1)$, and hence
	\begin{align}
		0<\mu<2
		\label{eq:mun_stab}
	\end{align}
	is sufficient for mean stability under \textbf{A1}--\textbf{A3}, for any waveform with finite augmented-regressor energy. Unlike the fixed-step bound \eqref{eq:mu_stab}, this admissible range is not set by $\lambda_{\max}(\mathbf{R})$. The normalization removes the step-size sensitivity to PAPR at the mean-stability level. However, waveform statistics still influence convergence through the conditioning of $\mathbf{R}$. Since the OEMP basis makes the covariance block $\mathbf{C}$ nearly block diagonal, the remaining conditioning effect is mainly associated with the pseudo-correlation block $\mathbf{P}$ that couples the direct and conjugate branches, as discussed next.
	
	\emph{Remark~1 ($\mathbf{P}=\mathbf{0}$ requires circularity beyond second-order properness).} The conditioning of the augmented correlation matrix $\mathbf{R}$ depends on the complementary-correlation matrix $\mathbf{P}=\mathbb{E}\{\boldsymbol{\psi}[n]\boldsymbol{\psi}^{T}[n]\}$, whose entries are higher-order pseudo-moments of the transmitted waveform. For the OEMP regressor, the zero-lag entries take the form $\mathbb{E}\{x^{q_1+q_2}|x|^{p_1+p_2-q_1-q_2}\}$ with $q_1+q_2\ge2$, whereas the off-diagonal entries involve the corresponding mixed pseudo-moments between delayed samples. Second-order properness, $\mathbb{E}\{x^2\}=0$, guarantees only that the entries with $q_1+q_2=2$ vanish. The remaining higher-order terms vanish only when $x$ is fully circular, i.e., when its distribution is invariant under arbitrary phase rotations, and remain small when the waveform is approximately circular. Consequently, when $\mathbf{P}\approx\mathbf{0}$, the augmented correlation matrix approximately decouples as $\mathbf{R}\approx\mathrm{diag}(\mathbf{C},\mathbf{C}^{*})$, yielding $\lambda_{\max}(\mathbf{R})\approx\lambda_{\max}(\mathbf{C})$ and $\mathrm{tr}(\mathbf{R})\approx2\,\mathrm{tr}(\mathbf{C})$. The circularity of the candidate waveforms is examined in Section~\ref{sec:waveforms} and verified numerically in Section~\ref{subsec:circularity_results}.
	
	For the widely linear canceler to identify a unique coefficient vector in the adopted basis space, $\mathbf{R}$ must be nonsingular. A necessary condition is that no basis function is perfectly non-circular, i.e., each $\psi_{q,p}[n]$ must satisfy $|\mathbb{E}\{\psi_{q,p}^{2}[n]\}|<\mathbb{E}\{|\psi_{q,p}[n]|^{2}\}$. This nonsingularity condition and the circularity condition in Remark~1 represent opposite limits of the same conditioning mechanism, and both affect the mean recursion in \eqref{eq:mean_error} through the eigenvalues of $\mathbf{R}$. The mean weight error is multiplied by $(\mathbf{I}-\mu\mathbf{R})$, so it decays in every direction only when all eigenvalues of $\mathbf{R}$ are strictly positive. At the perfectly non-circular limit, the direct and conjugate branches associated with some $\psi_{q,p}[n]$ become second-order collinear. Consequently, one eigenvalue of $\mathbf{R}$ reaches zero, leaving a nondecaying weight-error direction and an unidentifiable coefficient component. Circularity is the opposite limit. It drives the complementary block $\mathbf{P}$ toward zero, so $\mathbf{R}\to\mathrm{diag}(\mathbf{C},\mathbf{C}^{*})$ and the augmented recursion remains well conditioned, with $\lambda_{\min}(\mathbf{R})$ bounded away from zero and $\lambda_{\max}(\mathbf{R})\approx\lambda_{\max}(\mathbf{C})$ setting the fixed-step stability limit in \eqref{eq:mu_stab}. The magnitude of $\mathbf{P}$ therefore reflects both the conditioning of the augmented recursion and its distance from the singular limit. A waveform remains nonsingular as long as $|\mathbb{E}\{\psi_{q,p}^{2}[n]\}|$ stays below $\mathbb{E}\{|\psi_{q,p}[n]|^{2}\}$ for every basis function, which keeps $\lambda_{\min}(\mathbf{R})$ away from zero and allows the mean weight error to converge in every direction. The distance of the candidate waveforms from this limit is quantified by the impropriety measurements in Section~\ref{subsec:circularity_results}.
	
	\subsection{Steady-State EMSE and Covariance Evolution}
	\label{subsec:emse_misadj}
	Having characterized the mean behavior, we now examine the second-order performance. The quantity of interest is the EMSE, which determines the residual penalty introduced by finite-step-size adaptation and hence the attainable SIC floor. From \eqref{eq:residual_analysis}, the residual decomposes as
	\begin{align}
		d[n] = e_a[n] + \nu[n],
		\label{eq:residual_emse}
	\end{align}
	where
	\begin{align}
		e_a[n] \triangleq \underline{\tilde{\mathbf{w}}}^{T}[n]\underline{\boldsymbol{\psi}}[n] = \boldsymbol{\epsilon}^{H}[n]\underline{\boldsymbol{\psi}}[n]
		\label{eq:apriori_error}
	\end{align}
	is the a-priori error due to imperfect adaptation. The EMSE is
	\begin{align}
		\zeta[n]
		\triangleq
		\mathbb{E}\{|e_a[n]|^2\}
		=
		\mathrm{tr}\!\left(\mathbf{R}\mathbf{K}[n]\right),
		\label{eq:emse_def}
	\end{align}
	where
	\begin{align}
		\mathbf{K}[n]\triangleq \mathbb{E}\{\boldsymbol{\epsilon}[n]\boldsymbol{\epsilon}^{H}[n]\}
		\label{eq:K_def}
	\end{align}
	is the covariance matrix of the conjugate weight error.
	
	Forming the covariance recursion for $\mathbf{K}[n]$ from the weight-error recursion in \eqref{eq:weight_error_evol} and taking expectation under \textbf{A2}, the fixed-step realization $\mu[n]=\mu$ gives
	\begin{align}
		\mathbf{K}[n+1]={}&\mathbf{K}[n]-\mu\big(\mathbf{R}\mathbf{K}[n]+\mathbf{K}[n]\mathbf{R}\big)\nonumber\\
		&+\mu^2\big(\mathbf{R}\mathbf{K}[n]\mathbf{R}+\mathbf{R}\,\mathrm{tr}(\mathbf{R}\mathbf{K}[n])\big)+\mu^2\sigma_{\nu}^{2}\mathbf{R}.
		\label{eq:K_evolution}
	\end{align}
	Throughout this paper, $O(\mu^{k})$ denotes a quantity whose magnitude scales as $\mu^{k}$ for small step sizes rather than a computational cost. The $O(\mu)$ term $-\mu(\mathbf{R}\mathbf{K}[n]+\mathbf{K}[n]\mathbf{R})$ describes the mean-square contraction. The $O(\mu^2)$ terms have two distinct origins. The term $\mu^{2}\sigma_{\nu}^{2}\mathbf{R}$ is the disturbance injection, whereas $\mu^{2}\big(\mathbf{R}\mathbf{K}[n]\mathbf{R}+\mathbf{R}\,\mathrm{tr}(\mathbf{R}\mathbf{K}[n])\big)$ comes from the fourth-order regressor moment $\mathbb{E}\{\underline{\boldsymbol{\psi}}\underline{\boldsymbol{\psi}}^{H}\mathbf{K}\underline{\boldsymbol{\psi}}\underline{\boldsymbol{\psi}}^{H}\}$ generated by the quadratic part of the update. To obtain the closed form in \eqref{eq:K_evolution}, this fourth-order moment is evaluated using a circular-Gaussian reference model. Under this reference, Isserlis's theorem gives $\mathbb{E}\{\underline{\boldsymbol{\psi}}\underline{\boldsymbol{\psi}}^{H}\mathbf{K}\underline{\boldsymbol{\psi}}\underline{\boldsymbol{\psi}}^{H}\}=\mathbf{R}\mathbf{K}\mathbf{R}+\mathbf{R}\,\mathrm{tr}(\mathbf{R}\mathbf{K})$ \cite{SayedAdaptiveFilters2008}. This produces the two fourth-order terms in \eqref{eq:K_evolution}. Consistent with the cyclic-mean convention of Section~\ref{subsec:mean_stability}, the cyclostationary fourth moment is approximated by its stationary Isserlis form. The circular-Gaussian model is used only as a covariance-analysis reference, not as a model for the actual waveform statistics.
	This reference closure omits one additional contribution. The augmented regressor $\underline{\boldsymbol{\psi}}$ is improper by construction because its lower half is the conjugate of its upper half. Its pseudo-covariance is therefore $\widetilde{\mathbf{P}}=\mathbb{E}\{\underline{\boldsymbol{\psi}}\,\underline{\boldsymbol{\psi}}^{T}\}=\bigl[\begin{smallmatrix}\mathbf{P} & \mathbf{C}\\ \mathbf{C}^{*} & \mathbf{P}^{*}\end{smallmatrix}\bigr]$, which does not vanish even when $x$ is circular. Circularity removes only the diagonal blocks of $\widetilde{\mathbf{P}}$, while $\mathbf{C}$ remains in the off-diagonal blocks. A complete improper fourth-order closure would therefore add the complementary term $\widetilde{\mathbf{P}}\mathbf{K}^{*}\widetilde{\mathbf{P}}^{H}$ to \eqref{eq:K_evolution}. This term is also multiplied by $\mu^2$, so it is neglected together with the other $O(\mu^2)$ terms in the first-order steady-state analysis below.
	Equation \eqref{eq:K_evolution} serves as the circular-Gaussian reference recursion. Both the transient behavior and the steady-state floor are governed by the eigenvalue structure of $\mathbf{R}$, where a larger eigenvalue spread slows convergence and a larger trace raises the EMSE floor of the fixed-step realization.
	
	At steady state, setting $\mathbf{K}[n+1]=\mathbf{K}[n]=\mathbf{K}_{\infty}$ yields
	\begin{align}
		\mathbf{R}\mathbf{K}_{\infty}+\mathbf{K}_{\infty}\mathbf{R}
		={}&
		\mu\,\mathbf{R}\mathbf{K}_{\infty}\mathbf{R}
		+\mu\,\mathbf{R}\,\mathrm{tr}(\mathbf{R}\mathbf{K}_{\infty})
		+\mu\sigma_{\nu}^{2}\mathbf{R}.
		\label{eq:K_steady}
	\end{align}
	For small step sizes, $\mathbf{K}_{\infty}=O(\mu)$, so the right-hand-side terms carrying both $\mu$ and $\mathbf{K}_{\infty}$ are $O(\mu^2)$ and are dropped to first order, reducing \eqref{eq:K_steady} to
	\begin{align}
		\mathbf{R}\mathbf{K}_{\infty}+\mathbf{K}_{\infty}\mathbf{R}
		\approx
		\mu\sigma_{\nu}^{2}\mathbf{R}.
		\label{eq:K_steady_firstorder}
	\end{align}
	To solve \eqref{eq:K_steady_firstorder}, let $\mathbf{R}=\mathbf{U}\boldsymbol{\Lambda}\mathbf{U}^{H}$ be the eigenvalue decomposition of $\mathbf{R}$, where $\mathbf{U}$ is the unitary matrix of eigenvectors and $\boldsymbol{\Lambda}=\mathrm{diag}(\lambda_1,\ldots,\lambda_{2K})$ collects the corresponding eigenvalues, with $2K$ the augmented-regressor dimension. We further define $\bar{\mathbf{K}}_{\infty}\triangleq \mathbf{U}^{H}\mathbf{K}_{\infty}\mathbf{U}$. Substituting into \eqref{eq:K_steady_firstorder} and taking the $(i,j)$-th entry,
	\begin{align}
		(\lambda_i+\lambda_j)\,[\bar{\mathbf{K}}_{\infty}]_{ij}
		\approx
		\mu\sigma_{\nu}^{2}\lambda_i\,\delta_{ij},
		\label{eq:K_diag_entry}
	\end{align}
	where $\delta_{ij}$ is the Kronecker delta. The off-diagonal entries vanish to first order and the diagonal entries satisfy $[\bar{\mathbf{K}}_{\infty}]_{ii}\approx\tfrac{\mu}{2}\sigma_{\nu}^{2}$, so $\mathbf{K}_{\infty}\approx\tfrac{\mu}{2}\sigma_{\nu}^{2}\mathbf{I}$. Substituting into \eqref{eq:emse_def} gives
	\begin{align}
		\zeta_{\infty}
		\approx
		\frac{\mu}{2}\sigma_{\nu}^{2}\,\mathrm{tr}(\mathbf{R})
		\triangleq
		\zeta_{\infty}^{(\mathrm G)},
		\label{eq:emse_gauss}
	\end{align}
	which is the first-order circular-Gaussian reference floor. This reference is not needed to obtain the first-order floor itself. It is useful because it expresses the covariance recursion entirely in terms of $\mathbf{R}$, and it provides a baseline against which higher-order waveform effects can be isolated. The same first-order result can also be obtained from the energy-conservation method of \cite{SayedAdaptiveFilters2008}, which equates the mean weight-error energy before and after one update at steady state, without invoking a Gaussian regressor model. Taking the squared Euclidean norm of both sides of \eqref{eq:weight_error_evol}, imposing the steady-state condition $\mathbb{E}\{\|\boldsymbol{\epsilon}[n{+}1]\|^{2}\}=\mathbb{E}\{\|\boldsymbol{\epsilon}[n]\|^{2}\}$, and noting that the cross terms in $\nu[n]$ vanish under \textbf{A3}, the energy balance reduces to
	\begin{align}
		2\zeta = \mu\,\mathbb{E}\!\left\{\|\underline{\boldsymbol{\psi}}\|^2 \left|\underline{\boldsymbol{\psi}}^H\boldsymbol{\epsilon}\right|^2\right\} + \mu\sigma_\nu^2\,\mathrm{tr}(\mathbf{R}).
		\label{eq:energy_conservation_emse}
	\end{align}
	Since $\boldsymbol{\epsilon}=O(\sqrt{\mu})$ at steady state, the first term in \eqref{eq:energy_conservation_emse} is $O(\mu^2)$ and is the only term in this equality that depends on fourth-order regressor statistics. Neglecting it recovers \eqref{eq:emse_gauss} independently of the Gaussian approximation. The corresponding Gaussian-reference residual power is
	\begin{align}
		J_{\infty}^{(\mathrm G)}
		=
		\sigma_{\nu}^{2}+\zeta_{\infty}^{(\mathrm G)}
		\approx
		\sigma_{\nu}^{2}\left(1+\frac{\mu}{2}\mathrm{tr}(\mathbf{R})\right).
		\label{eq:J_infty_simple}
	\end{align}
	We next apply this first-order characterization to the proposed normalized update \eqref{eq:nlms}. For a compact first-order characterization, define the effective step
	\begin{align}
		\bar{\mu}
		\;\triangleq\;
		\frac{\mu}{\varepsilon_0+\mathrm{tr}(\mathbf{R})},
		\label{eq:mu_eff}
	\end{align}
	and adopt the standard NLMS separation approximation \cite{SayedAdaptiveFilters2008}:
	\begin{itemize}
		\item[\textbf{A4)}] On the $O(1/\bar{\mu})$ adaptation time scale, the augmented-regressor energy $\|\underline{\boldsymbol{\psi}}[n]\|^2$ varies much faster than the weight-error covariance. Hence, in first-order steady-state expressions, the data-dependent step $\mu[n]$ may be replaced by the effective step $\bar{\mu}$.
	\end{itemize}
	Under \textbf{A4}, the WL-NLMS recursion is approximated, to first order, by the fixed-step recursion with step size $\bar{\mu}$. Thus, $\bar{\mu}$ should be viewed as an analytical effective step, not as the exact average of the instantaneous step. By Jensen's inequality, $\mathbb{E}\{1/(\varepsilon_0+\|\underline{\boldsymbol{\psi}}\|^2)\}\ge 1/(\varepsilon_0+\mathrm{tr}(\mathbf{R}))$, so $\bar{\mu}$ slightly underestimates the mean instantaneous step, especially for heavy-tailed high-PAPR regressors. Substituting $\bar{\mu}$ for $\mu$ in \eqref{eq:emse_gauss} gives
	\begin{align}
		\zeta_{\infty}^{(\mathrm{NLMS})}
		\approx
		\frac{\bar{\mu}}{2}\sigma_\nu^2\,\mathrm{tr}(\mathbf{R})
		=
		\frac{\mu}{2}\sigma_\nu^2\,
		\frac{\mathrm{tr}(\mathbf{R})}{\varepsilon_0+\mathrm{tr}(\mathbf{R})}.
		\label{eq:emse_nlms_exact_eps}
	\end{align}
	For the small regularizer used only to avoid division by zero, $\varepsilon_0\ll\mathrm{tr}(\mathbf{R})$, this reduces to
	\begin{align}
		\zeta_{\infty}^{(\mathrm{NLMS})}
		\;\approx\;
		\frac{\mu}{2}\,\sigma_{\nu}^{2},
		\label{eq:emse_nlms}
	\end{align}
	which, unlike the fixed-step floor \eqref{eq:emse_gauss}, does not depend on $\mathrm{tr}(\mathbf{R})$ and is therefore waveform-independent to first order in $\mu$. This is the steady-state benefit of the proposed normalization. The first-order steady-state floor is now set only by the dimensionless step size and the uncancelable disturbance power, while the trace and eigenstructure of $\mathbf{R}$ remain waveform-dependent and govern only conditioning and convergence speed.
	
	When the regressors are non-Gaussian, fourth-order cumulants modify only the $\mu^2$ term in the covariance recursion, producing an $O(\mu^2)$ absolute correction to the EMSE, or an $O(\mu)$ relative correction to \eqref{eq:emse_gauss}, which is small at the step sizes used for deep SIC.
	
	\subsection{Amplitude Moments and Nonlinear SI Power}
	\label{subsec:papr_moments}
	
	For the EMP basis defined in \eqref{eq:emp_basis}, each raw component satisfies $|\xi_{q,p}[n]|=|x[n]|^{p}$, independent of $q$. Hence, the power of every branch of order $p$ is governed by the same envelope moment $\mathbb{E}\{|x[n]|^{2p}\}$ of the post-RRC signal. This motivates the normalized amplitude moment
	\begin{align}
		\gamma_p \;\triangleq\; \frac{\mathbb{E}\{|x[n]|^{2p}\}}{p!\,\sigma_x^{2p}},
		\qquad p=1,3,\ldots,P.
		\label{eq:gammap_def}
	\end{align}
	The normalizing factor $p!\,\sigma_x^{2p}$ equals $\mathbb{E}\{|x[n]|^{2p}\}$ for a circular complex Gaussian waveform, so $\gamma_p=1$ at every order in the Gaussian case, and $\gamma_p$ measures the departure of the $2p$-th envelope moment from that Gaussian value. Low-PAPR waveforms have more concentrated envelopes and hence $\gamma_p<1$ at high orders, whereas near-Gaussian high-PAPR multicarrier waveforms have $\gamma_p\approx1$. The separation grows with $p$, so the top-order moment $\gamma_P$ is the most sensitive indicator of waveform-dependent nonlinear SI.
	
	The second-order moment of each raw basis branch then follows as
	\begin{align}
		\mathbb{E}\{|\xi_{q,p}[n]|^2\} \;=\; \mathbb{E}\{|x[n]|^{2p}\} \;=\; \gamma_p\,p!\,\sigma_x^{2p},
		\label{eq:basis_moment}
	\end{align}
	so, under the common pulse shaping and power normalization used here, $\gamma_p$ is the dominant waveform-dependent factor in the diagonal energy of the raw basis Gram matrix.
	
\begin{figure}[!t]
		\centering
		\includegraphics[width=\columnwidth]{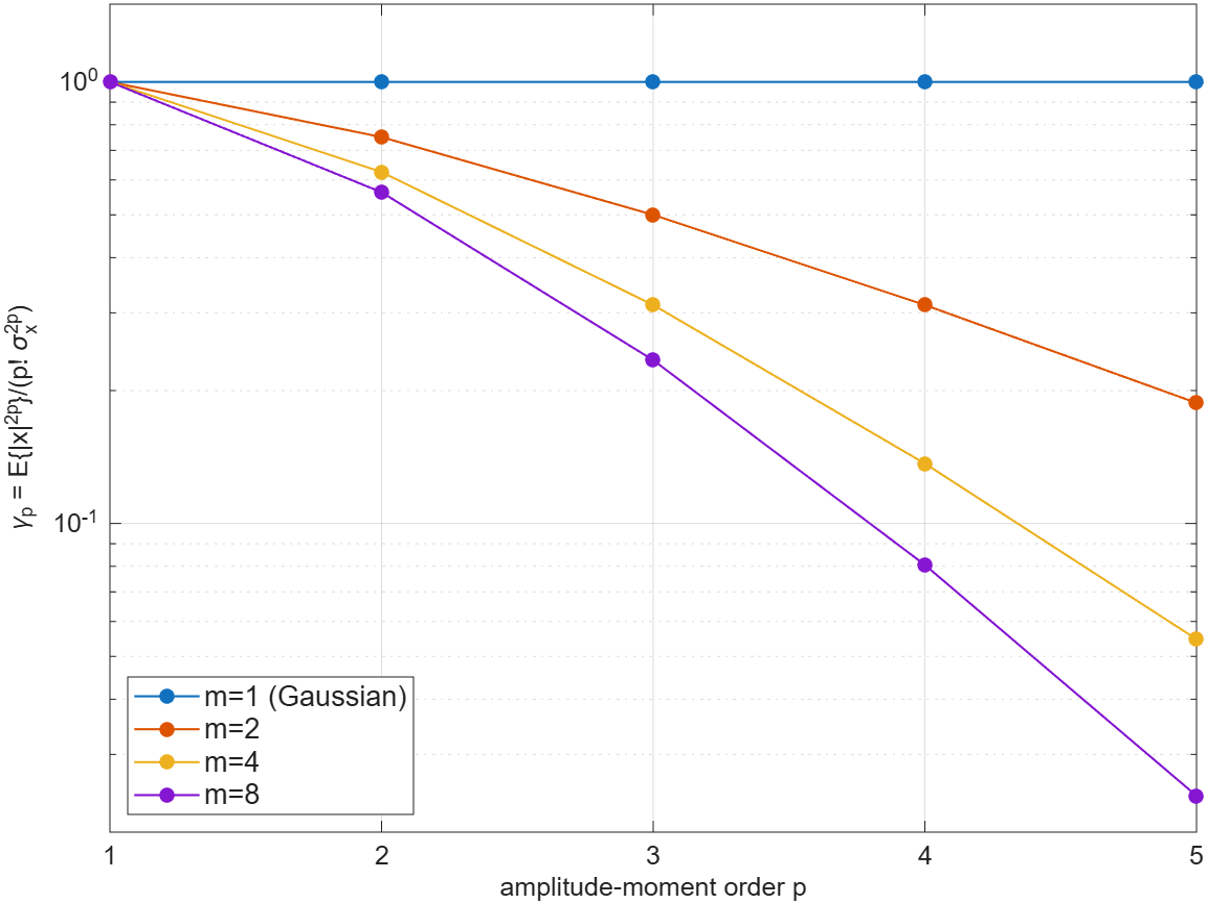}
		\caption{Normalized amplitude moment $\gamma_p$ versus order $p$ for a Gamma envelope-power model with shape $m$. The case $m=1$ (exponential power) is the circular-Gaussian anchor ($\gamma_p\equiv1$). Larger $m$ corresponds to a more concentrated, lower-PAPR envelope, and the moment gap widens with $p$.}
		\label{fig:gammap_theory}
	\end{figure}

	To illustrate how $\gamma_p$ reflects the envelope distribution, we use a Gamma model for the envelope power $|x[n]|^2$ with shape parameter $m$, which corresponds to a Nakagami-$m$ envelope \cite{Nakagami1960,SimonAlouini2005}. This model only serves as a one-parameter analytical proxy for the effect of envelope concentration, not as a fitted distribution for the four candidate waveforms. We have
	\begin{align}
		\gamma_p(m) = \frac{\Gamma(m{+}p)}{\Gamma(m)\,m^p\,p!},
		\label{eq:gammap_gamma}
	\end{align}
	where $\Gamma(\cdot)$ is the gamma function. The shape parameter $m$ controls the envelope concentration. When $m=1$, the model reduces to the exponential-power, circular-Gaussian anchor and gives $\gamma_p=1$. Larger values of $m$ produce more concentrated, lower-PAPR envelopes with smaller high-order moments. This separation widens rapidly with the nonlinear order, exceeding an order of magnitude by the fifth order for the more concentrated shapes, as shown in Fig.~\ref{fig:gammap_theory}. Thus, $\gamma_p$ provides an order-resolved measure of how the envelope distribution feeds the nonlinear PA orders.
	
	Combining \eqref{eq:basis_moment} with the PA model in \eqref{eq:pa_model}, the dominant waveform-dependent scaling of the SI power contributed by order $p$ can be written as
	\begin{align}
		P_p \;=\; |a_p|^2\,\gamma_p\,p!\,\sigma_x^{2p}, \qquad p=1,3,\ldots,P,
		\label{eq:Pp_decomp}
	\end{align}
	where $|a_p|^2=\sum_m|a_{p,m}|^2$ is the aggregate PA coefficient power at order $p$. This expression keeps only the dominant per-order amplitude scaling. Cross-terms between memory taps and basis branches are omitted because the pulse shaping, PA operating point, and cancellation basis are shared across the waveform candidates. Under common average-power normalization, $\gamma_1=1$, so the linear SI component does not distinguish the waveforms. The waveform-dependent burden enters through the nonlinear orders, whose powers scale with the corresponding $\gamma_p$ values. These same moments also affect the nonlinear blocks of the augmented correlation matrix $\mathbf{R}$. After the OEMP normalization removes the Gaussian factor $p!\,\sigma_x^{2p}$, the remaining order-dependent scaling is mainly captured by $\gamma_p$. Thus, high-PAPR waveforms can increase the nonlinear regressor energy and enlarge both $\lambda_{\max}(\mathbf{R})$ and $\mathrm{tr}(\mathbf{R})$. The proposed WL-NLMS normalization compensates for this first-order regressor-energy dependence in the adaptation step, but it does not change the nonlinear SI generated by the PA or the finite order and tap length of the cancellation basis. The residual waveform-dependent component is therefore tied to the part of the nonlinear SI that the adopted basis cannot represent.
	
	\section{Waveform Candidates and Their Suitability for SIC}
	\label{sec:waveforms}
	Section~\ref{subsec:papr_moments} identified complementary correlation and high-order amplitude moments as the waveform statistics most relevant to analog SIC. We now study the suitability of the four candidate waveforms under a common bandwidth and average-power normalization.
	
	\subsection{Candidate Waveforms}
	Each waveform generates the transmit baseband signal $x(t)$ that drives the PA model in \eqref{eq:pa_model}. Let $a_k$ denote zero-mean quadrature phase-shift keying (QPSK) symbols with $\mathbb{E}\{|a_k|^2\}=1$, and let $g(t)$ denote the unit-energy RRC pulse. The SC waveform is obtained by directly pulse shaping the QPSK symbols,
	\begin{align}
		x_{\mathrm{SC}}(t) = \sum_k a_k\, g(t-kT_s).
		\label{eq:sc_waveform}
	\end{align}
	Since SC does not superpose many independently modulated subcarriers, its envelope variation is mainly determined by RRC pulse shaping and oversampling. It therefore serves as the low-PAPR reference in the comparison.
	
	In contrast, OFDM forms each transmit block by applying an inverse DFT to the data symbols,
	\begin{align}
		\mathbf{x}_{\mathrm{OFDM}} = \mathbf{F}_{N}^{H}\mathbf{a},
		\label{eq:ofdm_waveform}
	\end{align}
	where $N$ is the number of subcarriers, $\mathbf{a}\in\mathbb{C}^{N}$ collects the QPSK symbols, and $\mathbf{F}_{N}$ denotes the $N$-point unitary DFT matrix. Since many independently modulated subcarriers are superposed, the time-domain samples approach a circular complex Gaussian distribution as the number of active subcarriers increases. This leads to the well-known high-PAPR envelope of OFDM \cite{HanLee2005,KorpiThesis2017}.
	
	OTFS places data symbols in the delay--Doppler domain and maps them to time-domain samples through time--frequency processing followed by cyclic-prefix OFDM (CP-OFDM) modulation. For the CP-OTFS implementation considered here,
	\begin{align}
		\mathbf{X}_{\mathrm{TF}} &= \mathbf{F}_{M_{\tau}} \mathbf{A} \mathbf{F}_{N_{\nu}}^{H}, \nonumber\\
		\mathbf{x}_{\mathrm{OTFS}}[:,n] &= \mathbf{F}_{M_{\tau}}^{H} \mathbf{X}_{\mathrm{TF}}[:,n],
		\label{eq:otfs_waveform}
	\end{align}
	where $\mathbf{A}\in\mathbb{C}^{M_{\tau}\times N_{\nu}}$ contains the delay--Doppler symbols \cite{HadaniOTFS2017,RavitejaOTFS2018}. Since the final transmission stage is CP-OFDM, the resulting time-domain samples inherit multicarrier-like envelope statistics.
	
	The final candidate is chirped DFT-s-OFDM, whose transmit block is
	\begin{align}
		\mathbf{x}_{\mathrm{cDFT}} = \boldsymbol{\Theta}\mathbf{F}_{N}^{H}\boldsymbol{\Pi}\mathbf{F}_{M}\mathbf{a},
		\label{eq:chirped_dfts_ofdm}
	\end{align}
	where $\mathbf{F}_{M}$ and $\mathbf{F}_{N}$ are unitary DFT matrices, $\boldsymbol{\Pi}$ maps the $M$ DFT-spread symbols onto the $N$ subcarriers, and $\boldsymbol{\Theta}=\mathrm{diag}\{c[n]\}$ applies the chirp $c[n]=e^{j\pi c_r n^2}$, in which $c_r$ is the chirp rate \cite{LiuTVT2025ChirpedDFTsOFDM}. DFT spreading gives the signal a single-carrier-like envelope and reduces PAPR relative to OFDM \cite{MyungSCFDMA2006}. The unit-modulus chirp changes only the phase and therefore preserves the instantaneous envelope.
	
	\subsection{Properness of the Candidate Waveforms}
	The above waveform constructions also clarify the role of second-order properness. For a zero-mean complex variable $z$, the impropriety coefficient is defined as $\kappa(z)\triangleq|\mathbb{E}\{z^2\}|/\mathbb{E}\{|z|^2\}\in[0,1]$, where $\kappa(z)=0$ corresponds to second-order properness \cite{SchreierScharf2010,NeeserTIT1993}. For a block vector, the corresponding properness condition is $\mathbb{E}\{\mathbf{x}\mathbf{x}^{T}\}=\mathbf{0}$. For any deterministic linear transform $\mathbf{T}$ acting on a proper data vector $\mathbf{a}$ with $\mathbb{E}\{\mathbf{a}\mathbf{a}^{T}\}=\mathbf{0}$,
	\begin{align}
		\mathbb{E}\{(\mathbf{T}\mathbf{a})(\mathbf{T}\mathbf{a})^{T}\} = \mathbf{T}\,\mathbb{E}\{\mathbf{a}\mathbf{a}^{T}\}\,\mathbf{T}^{T} = \mathbf{0}.
		\label{eq:proper_preserve}
	\end{align}
	Thus, the deterministic transforms used in OFDM, OTFS, and chirped DFT-s-OFDM preserve the second-order properness of the underlying QPSK or quadrature amplitude modulation (QAM) symbols. SC satisfies the same condition through linear pulse shaping, and the unit-modulus chirp in chirped DFT-s-OFDM only changes the sample phases. Therefore, within the waveform set considered here, second-order properness is shared by all candidates and does not distinguish their expected SIC behavior.
	
	\subsection{Envelope Statistics and SIC Prediction}
	\label{sec:compat}
	\label{subsec:predicted_ranking}
	With second-order properness common to all four candidates, the remaining waveform-dependent distinction comes from the envelope distribution of $x[n]$. Through the normalized amplitude moments $\gamma_p$, this distribution controls the nonlinear EMP components that enter the finite cancellation basis. Low-PAPR waveforms such as SC and chirped DFT-s-OFDM have more concentrated envelopes and should therefore produce smaller high-order moments. By contrast, the multicarrier time-domain samples of OFDM and CP-OTFS are closer to circular Gaussian and should produce moments closer to the Gaussian reference. Since the EMP basis functions scale as $|x[n]|^p$, this separation increases with the nonlinear order and is most pronounced at the highest retained order $P$.
	
	The resulting SIC floor can be interpreted by separating the residual power into three components,
	\begin{align}
		J_\infty = \sigma_{\nu,\mathrm{eff}}^2 + \zeta_\infty + P_{\mathrm{mis}},
		\label{eq:floor_decomp}
	\end{align}
	where $\sigma_{\nu,\mathrm{eff}}^2$ is the uncancelable disturbance power, corresponding to $\sigma_{\nu}^{2}$ in the analysis of Section~\ref{subsec:mean_stability}, $\zeta_\infty$ is the adaptation-induced EMSE floor, given for WL-NLMS by \eqref{eq:emse_nlms}, and $P_{\mathrm{mis}}$ is the residual nonlinear SI left outside the adopted finite-order, finite-tap basis, set by the basis order, tap length, PA operating point, and SI channel. Through the per-order SI powers in \eqref{eq:Pp_decomp}, $P_{\mathrm{mis}}$ scales with the normalized amplitude moments and is dominated by $\gamma_P$ at the highest retained order. Since \eqref{eq:emse_nlms} makes $\zeta_\infty$ waveform-independent to first order, the remaining waveform dependence should mainly appear through $P_{\mathrm{mis}}$.
	
	This leads to two SIC expectations under a common canceler configuration. First, SC and chirped DFT-s-OFDM should achieve lower residual SI than OFDM and OTFS because their smaller high-order moments reduce the nonlinear SI power exposed to finite-basis mismatch. Second, SC and chirped DFT-s-OFDM should behave similarly. DFT spreading gives chirped DFT-s-OFDM a single-carrier-like envelope, and the unit-modulus chirp preserves that envelope.
	
	\section{Simulation Results}
	\label{sec:sim_results}
	
	This section validates FASIC with WL-NLMS using SC, OFDM, OTFS, and chirped DFT-s-OFDM. All waveforms are evaluated with the same analog canceler, hardware impairments, passive suppression, and adaptation parameters, so that performance differences can be attributed to waveform statistics rather than waveform-specific tuning. The evaluation first verifies the relevant waveform statistics, then compares analog SIC depth, sensing preservation, and cancellation-span flexibility.
	
	\subsection{Simulation Setup}
	\label{subsec:sim_setup}
	
	The four candidate waveforms are evaluated under identical bandwidth and average-power constraints. All use QPSK modulation. Unless otherwise stated, the simulation parameters are listed in Table~\ref{tab:sys_params}. Each Monte Carlo realization contains $15\,000$ baseband symbols, and SIC depth, PAPR, and normalized amplitude moments are averaged over $N_{\mathrm{MC}}=50$ independent realizations.
	
	\begin{table}[!t]
		\centering
		\caption{Simulation System Parameters}
		\label{tab:sys_params}
		\setlength{\tabcolsep}{3pt}
		\resizebox{\columnwidth}{!}{%
			\begin{tabular}{ll}
				\toprule
				\textbf{Parameter} & \textbf{Value} \\
				\midrule
				Symbol period $T_s$ & 10 ns \\
				PA model & Parallel Hammerstein \\
				PA nonlinear order $P$ & 5 \\
				PA memory depth $M_{\mathrm{PA}}$ & 3 taps \\
				SI channel & 4-tap fractionally spaced \\
				Passive suppression & $-35$ dB \\
				canceler basis / update & OEMP / WL-NLMS \\
				Normalized step $\mu$ & 0.5 \\
				Adaptive filter taps $L$ & 24 \\
				Tap spacing & $T = T_s/2$ \\
				Baseband symbols / realization & 15000 \\
				Monte Carlo realizations $N_{\mathrm{MC}}$ & 50 \\
				Modulation & QPSK \\
				Sensing carrier / bandwidth & 2.4 GHz / 100 MHz \\
				Antenna gain (Tx, Rx) & 15 dBi \\
				OFDM subcarriers $N_{\mathrm{sc}}$ / CP & $64$ / $16$ \\
				DFT-s-OFDM spread, IFFT $(M,N)$ / CP & $(64,128)$ / $16$ \\
				DFT-s-OFDM chirp rate $c_r$ & $1/N$ \\
				OTFS delay bins $M_{\mathrm{otfs}}$ & 16 \\
				OTFS Doppler bins $N_{\mathrm{otfs}}$ & 16 \\
				OTFS frame size & 256 symbols \\
				OTFS implementation & CP-OTFS \cite{RavitejaOTFS2018} \\
				\bottomrule
		\end{tabular}}
	\end{table}
	
	The common RF and cancellation configuration is as follows. The transmitter uses the parallel-Hammerstein PA model in \eqref{eq:pa_model}, transmitter I/Q imbalance, a four-tap fractionally spaced SI channel, and $-35$~dB passive suppression. The analog canceler uses the OEMP basis in \eqref{eq:orth_basis_def} with the WL-NLMS update in \eqref{eq:nlms}. The normalized step size is $\mu=0.5$, and the adaptive filter has $L=24$ half-symbol-spaced taps for every waveform. No waveform-specific parameter tuning is used.
	For reproducibility, the transmitter I/Q imbalance consists of a $1$~dB amplitude mismatch and a $-5^{\circ}$ phase mismatch. Frequency-dependent I- and Q-branch responses are implemented by Butterworth filters whose bandwidths differ by $2\%$, producing a multi-tap conjugate response. The dominant PA coefficients are $a_{1,0}=1.05+j0.09$, $a_{3,0}=-0.05-j0.29$, and $a_{5,0}=-0.97-j0.70$, corresponding to an average transmit power of $20$~dBm. The SI channel has four fractionally spaced taps, $\{\tfrac{\sqrt2}{2}-\tfrac12 j,\,-0.4,\,0,\,0.3\}$, at delays $\{0,\,0.9,\,2.2,\,3.3\}\,T_s$, and is normalized to provide $-35$~dB passive suppression. Receiver noise is fixed at $60$~dB below the SI power. The sensing experiments use the radar configuration in Table~\ref{tab:sys_params}, with a $2.4$~GHz carrier, $15$~dBi antennas, and $100$~MHz bandwidth.
	
		\begin{figure}[!t]
		\centering
		\includegraphics[width=\columnwidth]{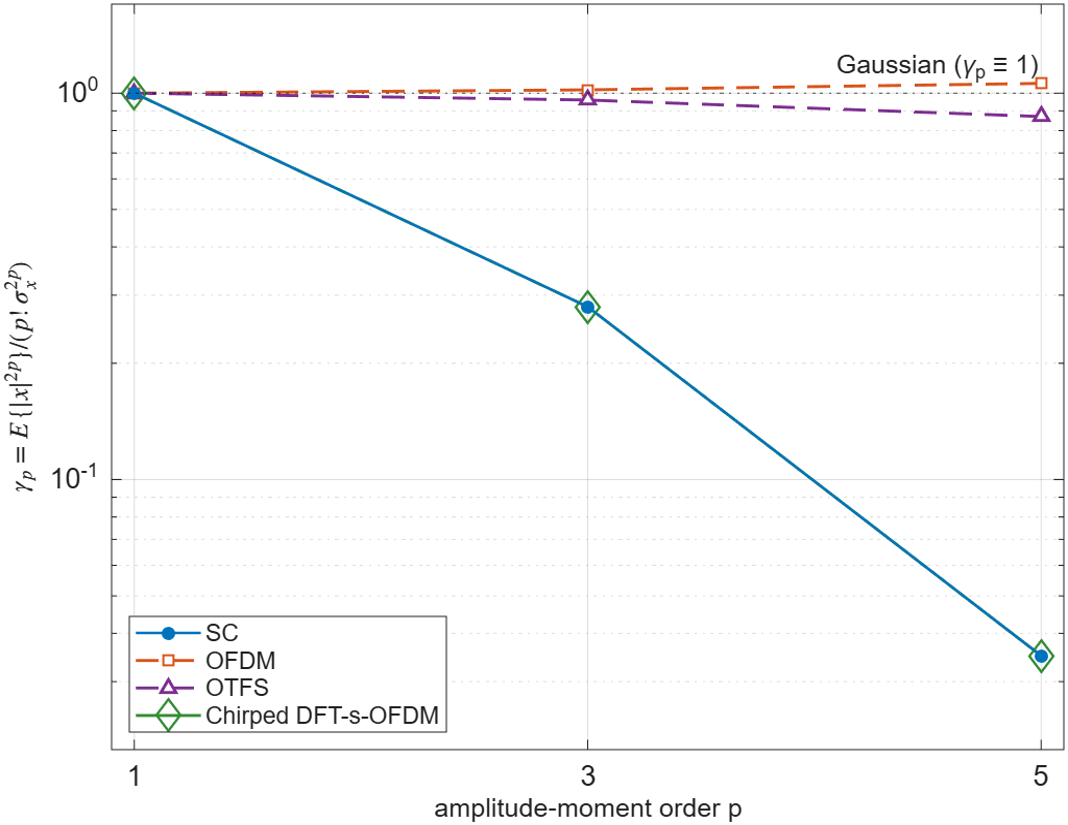}
		\caption{Measured normalized amplitude moment $\gamma_p$ versus order $p$ for the four waveforms. Measured counterpart of the envelope-power model in Fig.~\ref{fig:gammap_theory}.}
		\label{fig:gammap_measured}
	\end{figure}

	\subsection{Waveform Statistics}
	\label{subsec:circularity_results}
	
	The four waveforms separate into two envelope-statistics classes. SC and chirped DFT-s-OFDM have substantially lower PAPR than OFDM and OTFS, with a separation of approximately $4$~dB at the $99.9\%$ point of the complementary cumulative distribution function (CCDF), as summarized in Table~\ref{tab:waveform_results}. Chirped DFT-s-OFDM has nearly the same PAPR as SC. This agrees with the construction in Section~\ref{sec:waveforms}: DFT spreading produces a single-carrier-like envelope, and the unit-modulus chirp changes only the phase.
	
	\begin{table}[!t]
		\centering
		\caption{Waveform Comparison: PAPR and SIC Depth}
		\label{tab:waveform_results}
		\begin{tabular}{lcc}
			\toprule
			\textbf{Waveform} & \textbf{PAPR (dB)} & \textbf{SIC Depth (dB)} \\
			\midrule
			SC & $4.3$ & $33.9$ \\
			OFDM & $8.5$ & $30.3$ \\
			OTFS & $8.2$ & $30.9$ \\
			Chirped DFT-s-OFDM & $4.0$ & $33.9$ \\
			\bottomrule
		\end{tabular}
	\end{table}
	
	Fig.~\ref{fig:gammap_measured} reports the measured $\gamma_p$ at the odd PA orders $p=1,3,5$ of \eqref{eq:Pp_decomp}, computed on the post-RRC transmit signal of each waveform. OFDM and OTFS stay close to the Gaussian reference $\gamma_p\equiv1$, whereas SC and chirped DFT-s-OFDM fall progressively below it as the order grows, reaching $\gamma_5\approx0.035$ at the fifth order. At this order, $\gamma_5$ for the low-PAPR pair is smaller than for OFDM and OTFS by more than an order of magnitude, about $15$~dB, compared with roughly $4$~dB in PAPR.
	The order-resolved moments in \eqref{eq:gammap_def} therefore separate the waveforms more sharply than PAPR, which captures only the extreme tail of the envelope distribution. This confirms the mechanism identified in Section~\ref{subsec:papr_moments}. The nonlinear SI burden is governed by high-order envelope statistics, not by the peak amplitude alone.
	
	The measured differences are not caused by second-order impropriety. Consistent with \eqref{eq:proper_preserve}, all four waveforms are nearly proper at the waveform-generator output, with $\kappa(x)<0.05$. The deterministic IFFT, delay--Doppler, DFT-spreading, and chirp operations therefore preserve the properness of the underlying QPSK symbols. After the common RF front-end, the impropriety coefficient increases to approximately $\kappa\approx0.15$--$0.16$ because of the intentionally introduced transmitter I/Q imbalance.
	These measurements isolate the waveform-dependent factor relevant to analog SIC under the common RF configuration. The four candidates have nearly identical second-order properness after the shared hardware impairments, but their high-order envelope moments differ substantially. The normalized amplitude moments therefore serve as the primary statistical predictors of the nonlinear SI level and the residual analog SIC floor.
	
	\subsection{SIC Depth}
	\label{subsec:learning_sic}
	
	Table~\ref{tab:waveform_results} reports the SIC depths achieved by the four candidate waveforms. These depths form the two groups anticipated in Section~\ref{subsec:predicted_ranking}. SC and chirped DFT-s-OFDM reach approximately $33.9$~dB, whereas OFDM and OTFS reach $30.3$--$30.9$~dB. The two low-PAPR waveforms differ by less than $0.1$~dB, consistent with the single-carrier-like envelope produced by DFT spreading and the envelope-preserving unit-modulus chirp. Under the studied operating point, the gap between the low- and high-PAPR groups is approximately $3.3$~dB.
	
	All four waveforms reach these depths under a single FASIC configuration with no waveform-specific tuning, each converging reliably to more than $30$~dB of cancellation beyond the passive suppression. This confirms that the normalized update removes the fixed-step sensitivity to waveform-dependent regressor energy.
	The residual gap is attributed to finite-basis mismatch after the first-order adaptation floor has been equalized by WL-NLMS. OFDM and OTFS excite stronger high-order nonlinear PA components, leaving a larger portion of their nonlinear SI outside the adopted finite-order, finite-tap basis. In contrast, SC and chirped DFT-s-OFDM expose the canceler to weaker high-order nonlinear components, so their residual floors are closer to the uncancelable disturbance level.
	
	Together with the moment measurements in Section~\ref{subsec:circularity_results}, these SIC results show that the dominant waveform-dependent factor is the high-order envelope statistics, especially the highest retained moment, rather than the waveform family itself.

		\begin{figure*}[!t]
		\centering
		\subfloat[SC, before]{\includegraphics[width=0.24\textwidth]{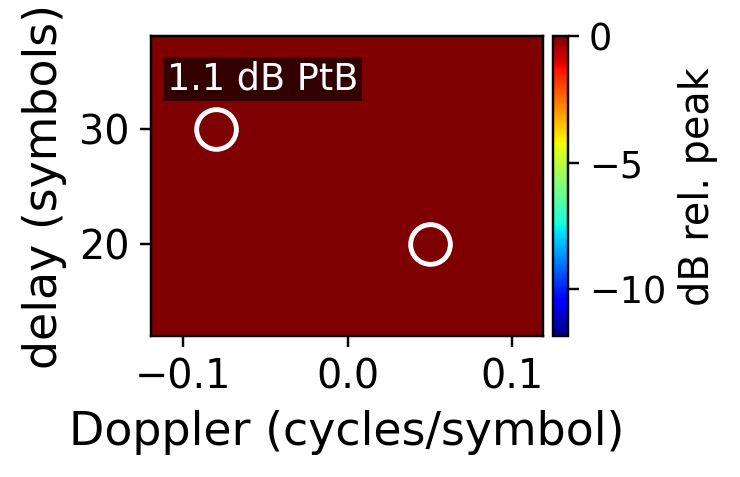}}\hfil
		\subfloat[OFDM, before]{\includegraphics[width=0.24\textwidth]{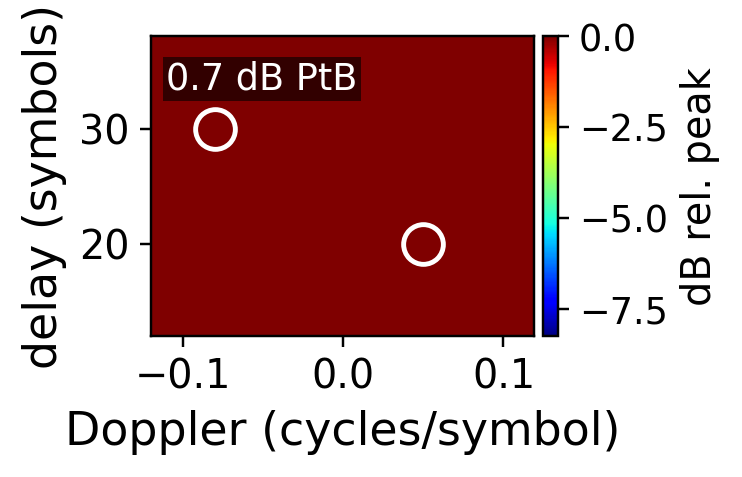}}\hfil
		\subfloat[OTFS, before]{\includegraphics[width=0.24\textwidth]{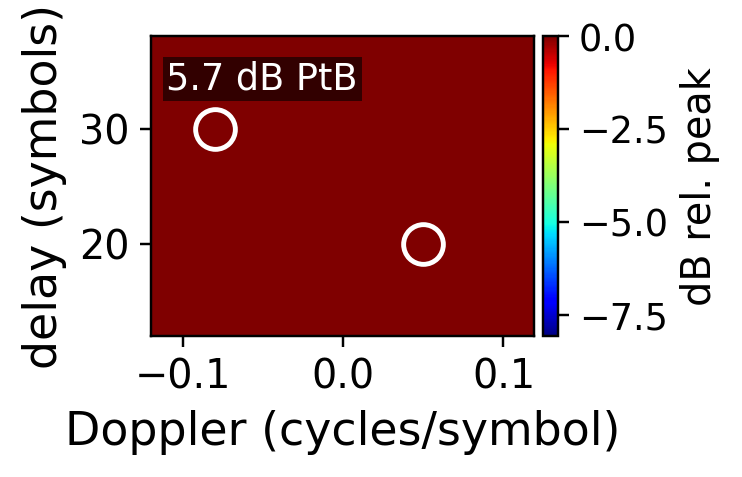}}\hfil
		\subfloat[Chirped, before]{\includegraphics[width=0.24\textwidth]{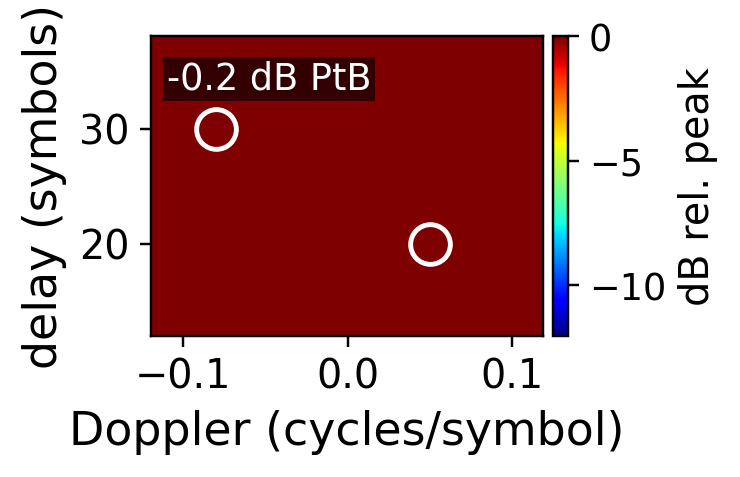}}\\
		\subfloat[SC, after]{\includegraphics[width=0.24\textwidth]{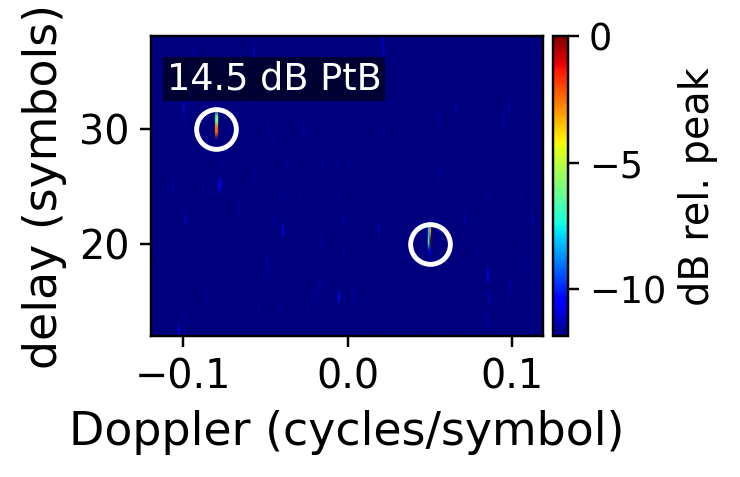}}\hfil
		\subfloat[OFDM, after]{\includegraphics[width=0.24\textwidth]{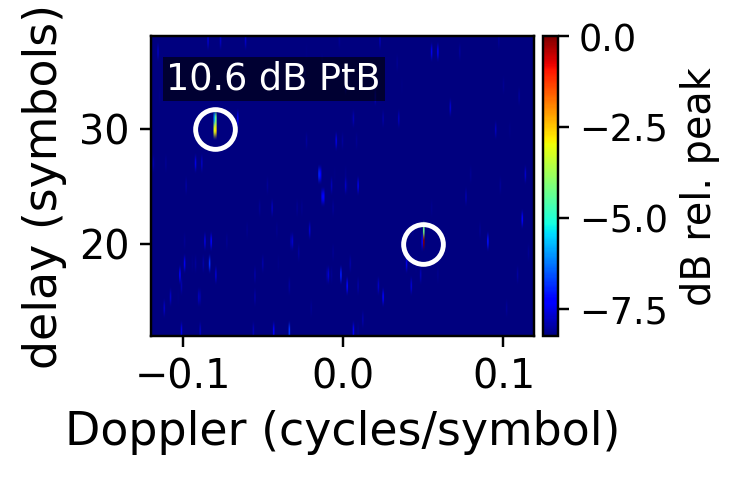}}\hfil
		\subfloat[OTFS, after]{\includegraphics[width=0.24\textwidth]{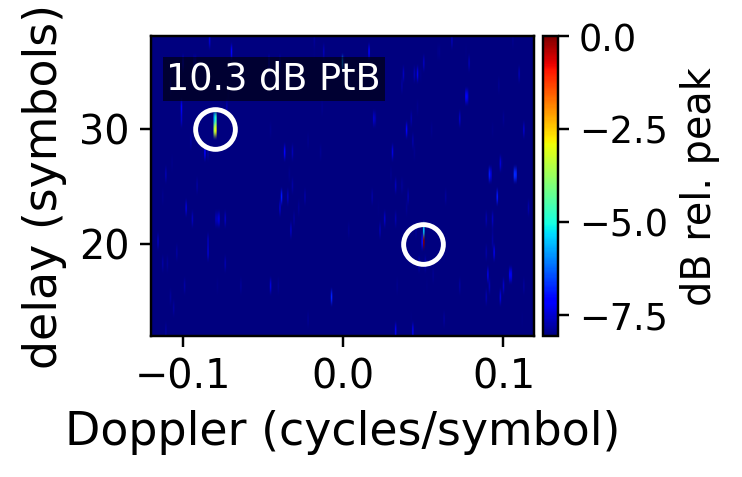}}\hfil
		\subfloat[Chirped, after]{\includegraphics[width=0.24\textwidth]{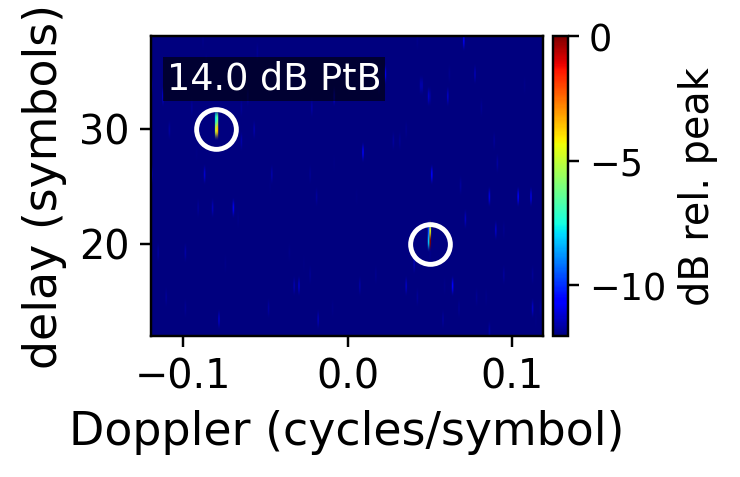}}
		\caption{Range-Doppler maps before (top row) and after (bottom row) FASIC for the four waveforms; true target cells are circled.}
		\label{fig:rd}
	\end{figure*}

	\begin{table}[!t]
		\centering
		\caption{Sensing on the post-FASIC residual: reflector CIR peak-to-floor ratio (before/after cancellation) and ToA error of the strongest reflector.}
		\label{tab:sensing}
		\setlength{\tabcolsep}{6pt}
		\footnotesize
		\begin{tabular}{lccc}
			\toprule
			\textbf{Waveform} & CIR before & CIR after & ToA err \\
			& [dB] & [dB] & [$T_s$] \\
			\midrule
			SC                 & $5.5$  & $21.7$ & $0$ \\
			OFDM               & $-0.7$ & $12.9$ & $0$ \\
			OTFS               & $4.8$  & $17.8$ & $0$ \\
			Chirped DFT-s-OFDM & $3.7$  & $19.9$ & $0$ \\
			\bottomrule
		\end{tabular}
	\end{table}

	\begin{figure}[!t]
		\centering
		\includegraphics[width=\columnwidth]{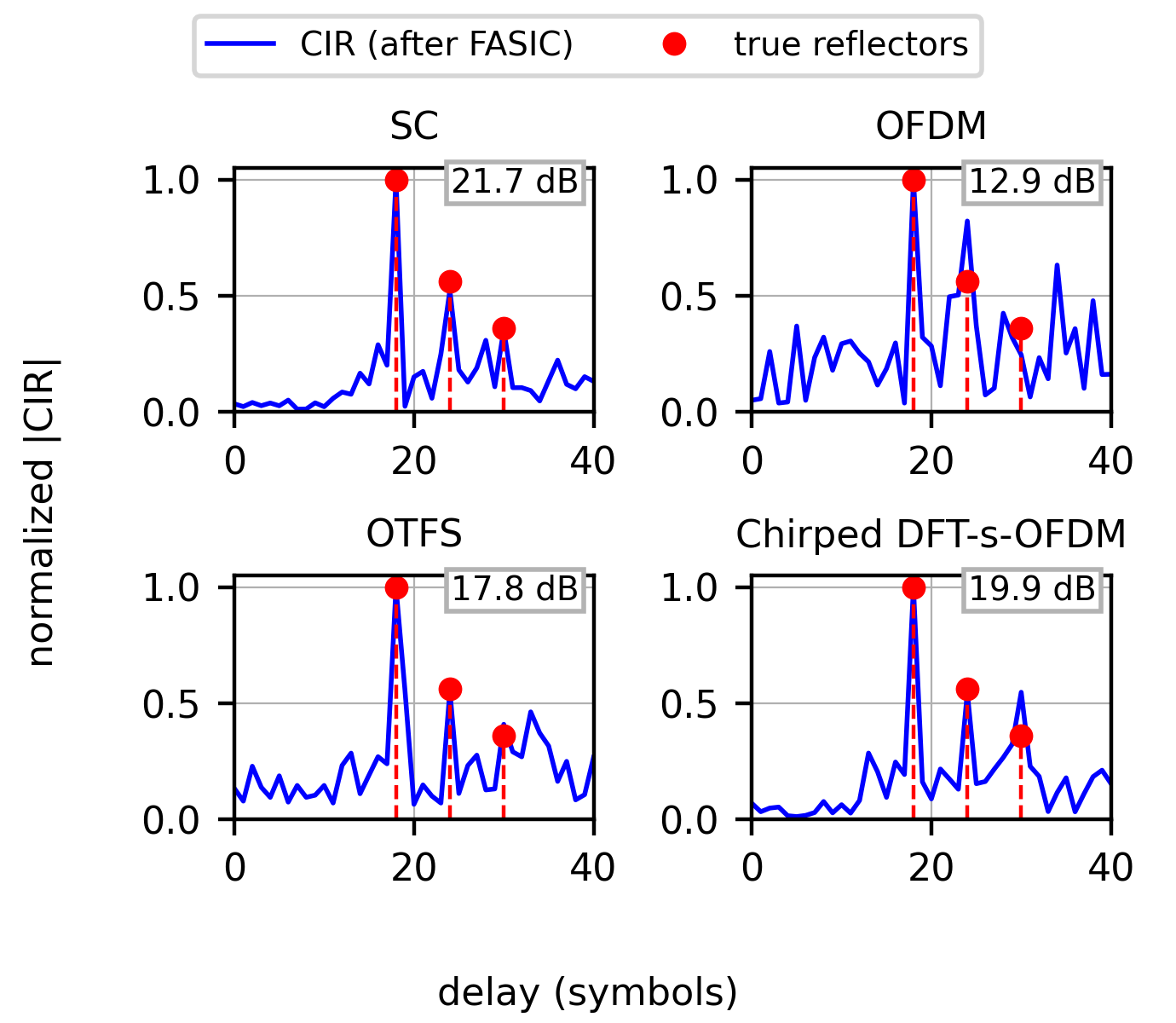}
		\caption{Recovered CIR/ToA for the four waveforms; the peak-to-floor margin is annotated in each panel.}
		\label{fig:cir}
	\end{figure}

	\subsection{Sensing Preservation}
	\label{subsec:sensing}
	A practical monostatic ISAC receiver must suppress strong SI without removing much weaker sensing echoes. We evaluate this requirement after FASIC has converged.
	
	Three static reflectors are placed at delays $\{18,24,30\}\,T_s$ (approximately $27$--$45$~m), which lie beyond the adaptive regressor span $(L{-}1)T=11.5\,T_s$. We then evaluate two sensing tasks on the post-cancellation residual. The first estimates the channel impulse response (CIR) and time of arrival (ToA) from a repeated-training preamble. The second reconstructs the range-Doppler map of two moving targets placed at delays $20$ and $30\,T_s$ with normalized Doppler shifts of $0.05$ and $-0.08$~cycles/symbol, respectively. These Doppler values are chosen for clear separation on the Doppler axis rather than to represent physical target speeds. The farther target has $2.5$ times the radar cross section of the nearer one but a net $3$~dB weaker echo. Both targets are evaluated after communication-payload removal by data wiping \cite{LiCSM2026,WangTCCN2026}.
	
	Table~\ref{tab:sensing} lists the CIR peak-to-floor ratios and ToA errors, and Fig.~\ref{fig:cir} shows the recovered CIRs for the four waveforms. The static-reflector experiment shows that FASIC preserves delayed sensing echoes for all four waveforms. After cancellation, the CIR peak-to-floor ratio increases to $12.9$--$21.7$~dB, compared with values at or below the residual floor before cancellation. The strongest reflector is recovered in the correct delay bin for every waveform, giving zero ToA error at the $1.5$~m range resolution. The largest margins occur for SC and chirped DFT-s-OFDM, with $21.7$ and $19.9$~dB, respectively. This follows the SIC-depth trend in Section~\ref{subsec:learning_sic}. A lower residual nonlinear SI floor leaves a cleaner sensing residual.
	
	Each range-Doppler map is formed by noncoherently averaging $N_{\mathrm{avg}}=8$ independent frames with a Hann slow-time window, and each before/after pair shares a color scale referenced to the after-FASIC peak. Fig.~\ref{fig:rd} shows the resulting maps. Before cancellation, the target responses are largely obscured by the dominant SI. After FASIC convergence, both moving targets are visible for every waveform, with peak-to-background ratios of approximately $10$--$15$~dB. SC and chirped DFT-s-OFDM give the cleanest maps.
	
	Together, the CIR/ToA and range-Doppler results show that FASIC preserves sensing information while improving the residual floor. Deeper analog cancellation increases the sensing margin, and the same cancellation architecture preserves both delay and Doppler information across all four waveform candidates.
	
		\begin{figure}[!t]
		\centering
		\includegraphics[width=\columnwidth]{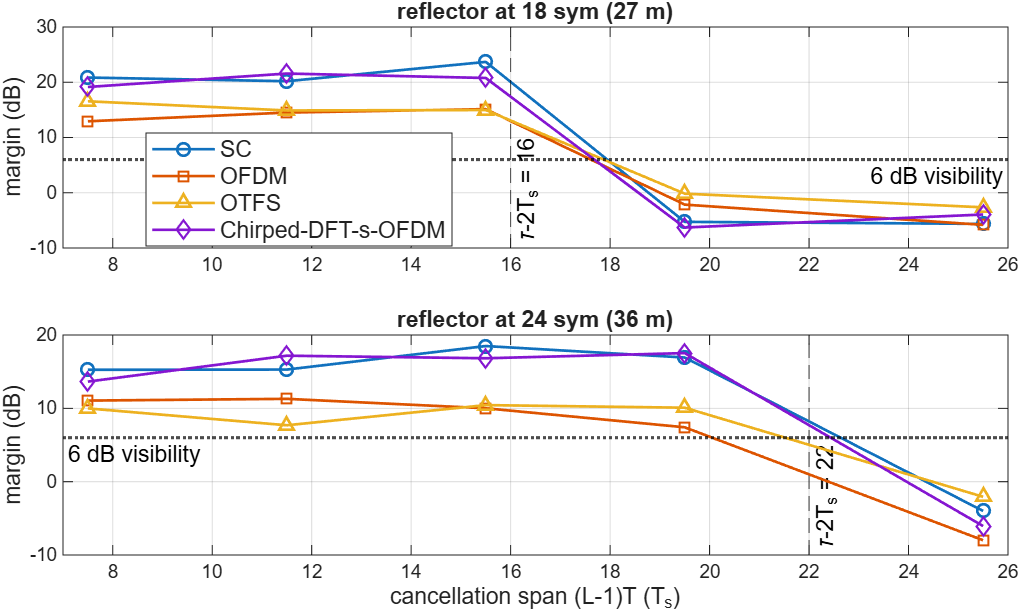}
		\caption{Per-reflector CIR margin versus cancellation span for the four waveforms. The $6$~dB visibility threshold is shown by the dotted line, and the predicted delay boundaries $\tau-2\,T_s$ are marked.}
		\label{fig:span_margin}
	\end{figure}

	\subsection{Cancellation-Span Flexibility}
	\label{subsec:span_flex}
	
	To validate the delay-span condition in \eqref{eq:span_gate}, we vary the adaptive filter length $L$ so that the cancellation span successively reaches the two static reflectors at $18\,T_s$ and $24\,T_s$. Visibility is measured by the CIR margin, defined as the recovered peak above the local residual floor. A reflector is considered visible when this margin exceeds $6$~dB.

	Fig.~\ref{fig:span_margin} plots the per-reflector CIR margin versus the cancellation span for the four waveforms. Each reflector remains visible while its delay lies outside the effective cancellation span. Once the span crosses the delay boundary $\tau-2\,T_s$, the corresponding margin drops below the visibility threshold and approaches the residual floor. For all four waveforms, this transition occurs within one delay-grid interval of the predicted boundary. At an intermediate span, the $18\,T_s$ reflector is absorbed while the $24\,T_s$ reflector remains visible. The SIC depth remains essentially unchanged once the span covers the SI delay spread, and degrades only when the span is too short to represent the SI channel.
	
	These results demonstrate delay-selective suppression of static reflectors. Static echoes whose delays fall inside the selected cancellation span are absorbed by the adaptive canceler, whereas delayed echoes outside the span are preserved. Consistent with the decorrelation mechanism, moving targets at the Doppler frequencies considered here remain visible even when their delays are shorter than the cancellation span.

	\section{Conclusion}
	This paper developed FASIC with WL-NLMS as a waveform-robust analog SIC solution for monostatic ISAC. It combines an OEMP basis, widely linear reconstruction, and normalized adaptation to suppress direct, conjugate, and nonlinear SI before the receiver front-end. Its configurable cancellation span further provides delay-selective suppression of near-field static clutter.
	
	The analysis separated two waveform-dependent mechanisms namely second-order impropriety and high-order amplitude moments. Second-order impropriety affects the conditioning of the widely linear recursion and is accommodated by the augmented direct--conjugate structure. The high-order amplitude moments determine the nonlinear SI power presented to the finite cancellation basis, and therefore govern the residual SIC floor after WL-NLMS equalizes the first-order adaptation floor.
	
	Simulations under one common hardware and canceler configuration supported the analysis. FASIC achieved stable analog SIC of approximately $30$--$34$~dB for SC, OFDM, OTFS, and chirped DFT-s-OFDM without waveform-specific tuning. The low-PAPR waveforms achieved an approximately $3.3$~dB deeper floor than OFDM and OTFS. Their smaller high-order amplitude moments reduce the nonlinear SI presented to the finite cancellation basis. The post-cancellation residual supported CIR/ToA estimation and range-Doppler imaging for all four waveforms, while the cancellation-span experiment illustrated how static reflectors can be absorbed or preserved according to the selected delay span. 
    Future work will extend FASIC to frequency-selective and time-varying SI channels, investigate richer nonlinear basis representations and adaptive normalization strategies for high-PAPR waveforms, and validate the design using hardware prototypes.
	
	\bibliographystyle{IEEEtran}
	\bibliography{refs}
	
\end{document}